\documentclass[aps,prx,reprint,amsmath,amssymb,amsfonts,longbibliography]{revtex4-2}

\usepackage[english]{babel}
\usepackage[utf8x]{inputenc}
\usepackage[T1]{fontenc}

\usepackage[a4paper,top=3cm,bottom=2cm,margin = 25mm]{geometry}
\usepackage{bbold}

\usepackage{amsmath}
\usepackage{amssymb}
\usepackage{graphicx}
\usepackage{verbatim}
\usepackage{booktabs}
\usepackage{hyperref}
\usepackage{xcolor}

\newcommand{\ket}[1]{\left| #1 \right\rangle} 
\newcommand{\bra}[1]{\left\langle #1 \right|} 
\newcommand{\innProd}[2]{\left\langle #1 \middle| #2 \right\rangle} 

\newcommand{\matel}[2]{
\ifnum 0=#1\relax
S_{#2}
\else
S_{#2}^{*}
\fi
}
\newcommand{\matrixelement}[3]{\langle #1 | #2 | #3\rangle}

\begin{document}
\title{Photoelectron signature of dressed-atom stabilization in intense XUV field}
\author{Edvin Olofsson}
\affiliation{Department of Physics, Lund University, Box 118, SE-221 00 Lund, Sweden}



\author{Jan Marcus Dahlström}
\email{marcus.dahlstrom@matfys.lth.se}
\affiliation{Department of Physics, Lund University, Box 118, SE-221 00 Lund, Sweden}

\begin{abstract}
    Non-perturbative resonant multiphoton ionization $(1+1)$ is studied using the resolvent operator technique. Scaling parameters for effective two-level Hamiltonians are computed for hydrogen and helium atoms to provide a quantitative description of Rabi oscillations at XUV wavelengths, which were recently observed using a seeded Free-Electron Laser [S. Nandi  \textit{et al.},   Nature \textbf{608}, 488-493 (2022)]. The resulting photoelectron spectra exhibit a range of Autler-Townes doublets, which are studied for different intensities, detunings and interaction times. We identify a photoelectron signature that originates from stabilization against ionization of helium atoms interacting with intense circularly polarized XUV light. Thus, our work shows how it is possible to test the prediction of dressed-atom stabilization by Beers and Armstrong [B. L. Beers and L. Armstrong, Phys. Rev. A \textbf{12}, 2447 (1975)], without the demanding requirement of atomic saturation in the time domain.  
\end{abstract}

\maketitle

\section{Introduction}\label{sec:Introduction}
Free-electron laser (FEL) facilities around the world provide intense radiation in the extreme ultraviolet (XUV) and X-ray regimes~\cite{huang_features_2021}, which has opened up for studies of non-linear light--matter interaction at short wavelengths in atoms and molecules~\cite{ackermann_operation_2007,emma_first_2010,ishikawa_compact_2012,allaria_highly_2012,mirian_generation_2021,young_femtosecond_2010,rudenko_femtosecond_2017,young_roadmap_2018,lindroth_challenges_2019}. Recent advances include applications of isolated attosecond pulses~\cite{duris_tunable_2020,li_attosecond_2022} and attosecond pulse trains~\cite{maroju_attosecond_2020,maroju_attosecond_2023} from FELs. 
While high-order harmonic generation (HHG) can be used to produce highly coherent pulses in the XUV and soft X-ray ranges \cite{lewenstein_theory_1994}, the intensity from such table-top sources is too low to drive non-linear processes.  
In contrast, FEL sources based on self-amplified spontaneous emission (SASE) can reach high intensities and photon energies, but the coherence properties of the pulses are limited~\cite{huang_features_2021}.  For this reason, laser-seeded FELs present some attractive properties that include high intensity, when compared with isolated harmonics from HHG sources, and good coherence properties and shot-to-shot reproducibility, when compared with SASE-FEL sources~\cite{allaria_highly_2012,huang_features_2021}.
Recently, these properties allowed for observation of Rabi dynamics at XUV wavelengths in helium atoms using a two-photon resonant photoionization processes, in an experiment performed by Nandi \textit{et al.} using the seeded FEL at FERMI~\cite{Nandi2022}. 

Rabi oscillations are signatures of non-linear coherent quantum dynamics that appear when a system is well described as an interacting two-level system~\cite{Rabi1937,Autler1955}. Autler and Townes showed that a consequence of Rabi oscillations is the spectroscopic splitting of an absorption line into two lines, the so-called Autler-Townes (AT) doublet \cite{Autler1955}. The two spectral components of the doublet are separated by the generalized Rabi frequency, $W=\sqrt{\Omega^2+\Delta\omega^2}$. 
Short-wavelength FELs have been the inspiration for several theoretical papers exploring resonant multiphoton ionization both through models~\cite{rodriguez_resonant-enhanced_2009,Toth2021,Younis2022,Bunjac2022,Zhang2022} and by numerically solving the time-dependent Schrödinger equation (TDSE)~\cite{LaGattuta1993,Girju2007,Toth2021,Younis2022,Zhang2022}.
In seminal works, Beers and Armstrong \cite{Beers1975}, Knight~\cite{knight_saturation_1977}, and Holt, Raymer and Reinhardt \cite{Holt1983} used an effective two-level model to study resonant multiphoton ionization of atoms, taking into account the interaction with the rest of the Hilbert space through effective parameters entering the Hamiltonian of the two-level system~\cite{CT1998}. It was shown that the ionization can not in general be described by a model with a simple exponential decay. The total amount of ionization has an intricate dependence on the parameters of the effective Hamiltonian and the interaction time, with several different parameter regimes identified \cite{Beers1975,Holt1983}. It was also predicted that an AT-splitting should appear in the photoelectron spectrum for sufficiently intense pulses~\cite{knight_saturation_1977}. 
In the special case of a {\it single} ionization continuum, it was predicted that interference between resonant and non-resonant ionization processes could lead to incomplete ionization \cite{Beers1975}, which we interpret as a form of {\it stabilization} against photoionization from the atom in a strong field. 
Interestingly, the recent experiment by Nandi \textit{et al.} was performed at an intensity where the rates of the resonant and non-resonant processes are comparable~\cite{Nandi2022}, but the continuum consisted of two parts ($s$- and $d$-wave). This opens the question if stabilization against photoionization is possible in this more general case and if such effects can be inferred from energy-resolved photoelectron distributions.

In this work we explore an extension of the model used in Nandi \textit{et al.} \cite{Nandi2022}, taking into account the effects of AC-Stark shifts, complex Rabi frequencies and depletion through an effective Hamiltonian describing the resonant two-level system,  following the the works \cite{Beers1975,Holt1983}. 
%
The article is organized as follows. 
In Sec.~\ref{sec:theory} we present the theoretical foundation of our work. In Sec.~\ref{sec:resolvent} the resolvent operator technique is reviewed and applied to the particular case of an effective two-level system. In Sec.~\ref{sec:param} the parameters for couplings in hydrogen $1s-2p$ and helium $1s^2-1s2p$ are computed by perturbation theory and then contrasted with known values from literature (see Appendix~\ref{app:comparison}). In Sec.~\ref{sec:rates} and \ref{sec:amplitudes} photoionization rates and energy-resolved photoelectron distributions are extracted from the effective two-level system. Our results for two-photon resonant ionization of helium atoms via the $1s^2-1s2p$ transition, for both circularly (single) and linearly  (double) FEL pulses (photoelectron continua), are shown in Sec.~\ref{sec:results} (see also Appendix~\ref{app:beta}). Conclusions are presented in Sec.~\ref{sec:conclusions}. Atomic units are used unless otherwise stated: $\hbar=e=4\pi\epsilon_0=m_e=1$.


\section{Theory}\label{sec:theory}

There have been several studies of energy-resolved photoelectron spectra from resonant multiphoton ionization, starting with pioneering numerical studies on resonant two-photon ionization of hydrogen atoms that revealed AT doublets in the above-threshold ionization (ATI)~\cite{LaGattuta1993,Girju2007}, and in the photoelectron spectra of ionized H$_2$ molecules~\cite{palacios_step-ladder_2006}. In Ref.~\cite{Girju2007} attempts were made to model the photoelectrons of a Rabi cycling atom using the Strong Field Approximation (SFA), but this failed to capture some aspects of the results from TDSE simulations.  In the same work AC-stark shifts and ionization rates were estimated from Floquet theory, which provide information about the dynamics of the atom in the field, but its connection to a time-dependent picture is not straightforward \cite{Holt1983,Girju2007}. 
%
%
Models including envelope effects have been studied in both two-photon~\cite{Demekhin2012,Younis2022,Bunjac2022,Zhang2022} and three-photon~\cite{Toth2021} resonant photonionization, that also include the effect of AC-Stark shifts. An asymmetry of the AT doublet with respect to detuning of the FEL field was predicted in Refs.~\cite{LaGattuta1993,Girju2007}, and later observed experimentally Ref.~\cite{Nandi2022}. Using theoretical models, it has been shown that the asymmetry of the AT doublet depends on the relation between resonant and non-resonant pathways and that the energetic condition of the symmetric AT doublet is sensitive to associated quantum interference effects and FEL pulse shape  ~\cite{Nandi2022,Zhang2022}. 

\begin{figure*}
\includegraphics[width=0.9\textwidth]{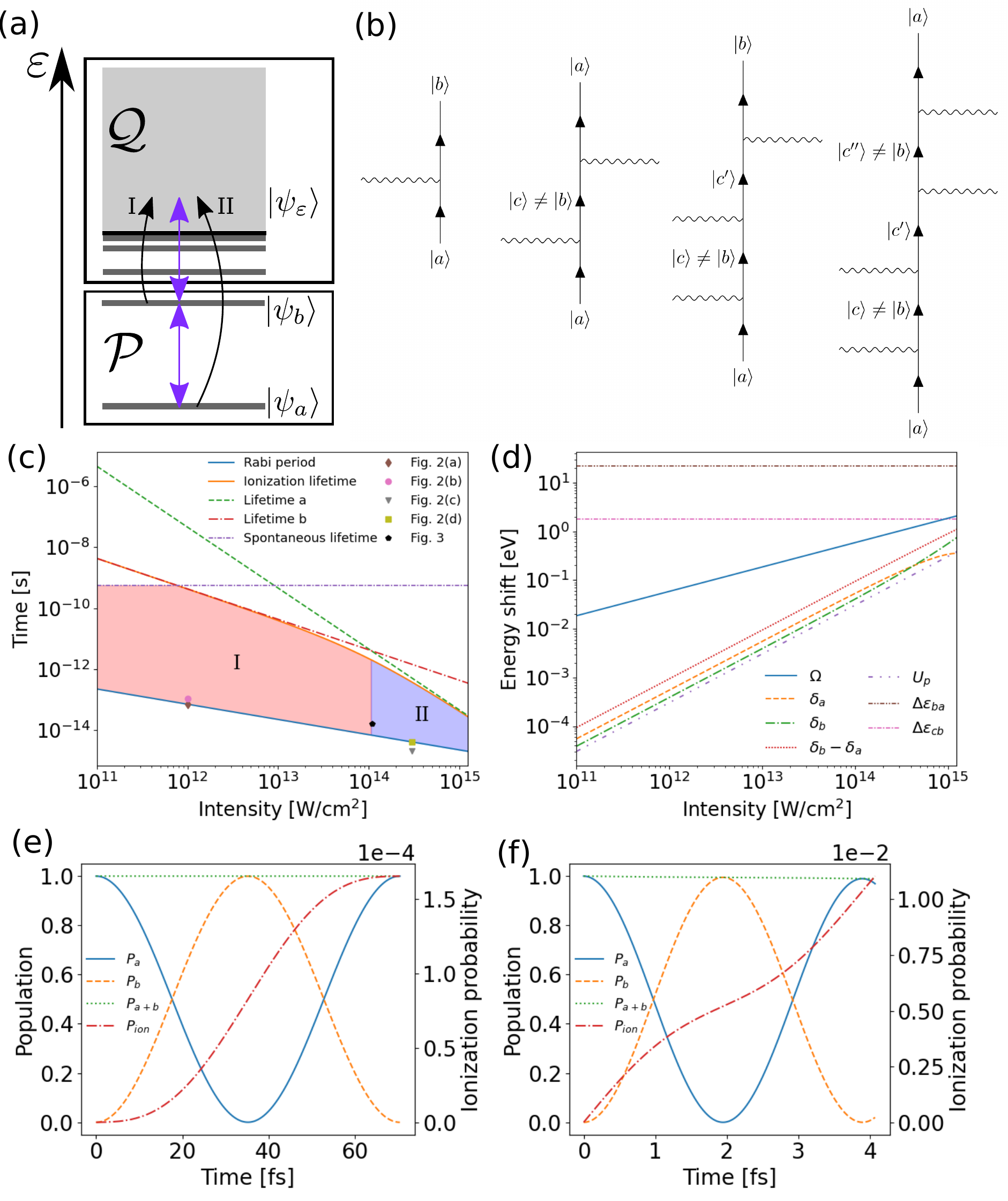}
\caption{\label{fig:levels} Panel (a) shows a schematic atomic level diagram of the situation under consideration, with the subspaces $\mathcal{P}$ and $\mathcal{Q}$ indicated. Panel (b) shows examples of the processes starting in state $\ket{a}$ considered in each order of perturbation theory. The processes starting from $\ket{b}$, and the other time-orderings can be constructed analogously. Absorption and emission of a photon is indicated by a wavy line on the left and right sides of the diagram, respectively. The resonant intermediate states are excluded from the intermediate state sums by projection operators. (c) Map of relevant timescales for a He atom undergoing Rabi oscillations to the $1s2p$ state. The colored regions denote the domains where the resonant (I) or non-resonant (II) processes dominate. (d) Rabi frequency $\Omega$, AC-Stark shifts of the states in $\mathcal{P}$ due to interaction with the states in $\mathcal{Q}$, and ponderomotive shift $U_p$ as a function of FEL intensity. The field-free resonance energy $\Delta \epsilon_{ba}$ and the distance $\Delta \epsilon_{cb}$ to the non-resonant state closest in energy to $b$ are shown as horizontal lines. (e) and (f) Populations in the two-level system and ionization probability for one Rabi period at an intensity of $1\times 10^{12}$ W/cm$^2$ (e) or $3\times 10^{14}$ W/cm$^2$ (f). }
\end{figure*}

In this work we make use of the {\it resolvent operator} formalism, which was used by Beers and Armstrong \cite{Beers1975} for resonant multiphoton ionization of atoms. The main advantage of this approach is that the resulting model for an effective two-level system is fully analytical and that its parameters can be unambigously determined using perturbation theory \cite{CT1998}. The problem is formulated using quantum optics with the total ``coupled'' Hamiltonian 
\begin{equation}
H=H_A+H_F+H_{AF},
\label{eq:H_QO}
\end{equation}
where the ``uncoupled'' atom--field Hamiltonian: $H_0=H_A+H_F$ is the sum of the atomic $H_A$ and field $H_F$ parts, while $V=H_{AF}$ describes the coupling of the atom with the field. We consider an atom in its ground state, $\psi_a$, with a single mode of radiation in a Fock state with $N_0$ photons of frequency $\omega_0$, which is an eigenstate of the uncoupled atom-field Hamiltonian: $H_0\ket{a}=E_a\ket{a}$, where $E_a=\epsilon_a+N_0\omega_0$ and $\ket{a}=\ket{\psi_a,N_0}$. We assume that there is one excited atomic state: $\psi_b$, that is strongly coupled through absorption of one photon by the atom:  $\ket{b}=\ket{\psi_b,N_0-1}$ with $E_b\approx E_a$, as shown in Fig.~\ref{fig:levels}~(a). The strength of the coupling is determined by half the Rabi frequency, $\matrixelement{b}{V}{a}=\Omega_0/2=z_{ba}{\cal E}_0/2$, where $z_{ba}$ is the atomic dipole matrix element and ${\cal E}_0$ is the electric field strength in the semi-classical approximation. The detuning of the interaction, denoted $\Delta\omega=E_{ab} = \omega_0-(\epsilon_b-\epsilon_a)$ with $E_{ab}=E_a-E_b$, is assumed to be small compared to other bound transitions: $|\Delta\omega|<|E_{ci}|$ with $i\in\{a,b\}$, which implies that $V$ is ``weak'' outside the strongly coupled two-level subspace: $\mathcal{P}$ spanned by: $\ket{a}$ and $\ket{b}$, see Fig.~\ref{fig:levels} (a). The orthogonal complement $\mathcal{Q} = \mathcal{P}^{\perp}$ is spanned by complement atomic and field states: $\ket{c}$. In this work we will consider up to two exchanged photons: $|\Delta N|\le 2$, as shown in Fig.~\ref{fig:levels}~(b). Net absorption of two photons, $N_0\rightarrow N_0-2$, leads to ionization into the photoelectron continuum with angular momentum: $\ell$, and kinetic energy $\epsilon_\mathrm{kin}\approx 2\omega_0-I_p$, where $I_p$ is the binding potential of the atom. 

Effects beyond the rotating-wave approximation (RWA) are incorporated in the model, such as the Bloch-Siegert shift of the resonance, AC Stark-shifts due to other states and consistent photoionization from the two-level system including quantum interference effects.  
All that is required is an {\it effective} two-level Hamiltonian: 
\begin{equation}
    H_{\textrm{eff}} =  
    \begin{bmatrix}
     h_{aa} & h_{ab}\\
     h_{ba} & h_{bb}
    \end{bmatrix},
    \label{eq:Heff}
\end{equation}
which is non-Hermitian with complex matrix elements: $h_{aa},\, h_{bb}, \, h_{ab} =  h_{ba} \in \mathbb{C}$. This type of effective models is sometimes considered as phenomenological, c.f. \cite{Girju2007} with intensity dependent parameters estimated by Floquet theory \cite{Holt1983} or by comparison with TDSE simulations \cite{Zhang2022}. Here, however, we compute the parameters using level-shift scaling factors from first-principle perturbative expansions \cite{CT1998}. Calculations are performed for hydrogen and helium, at the level of configuration-interaction singles (CIS) \cite{greenmanImplementationPRA2010}, with a numerical technique called exterior complex scaling (ECS) to take into account integration over intermediate continuum states \cite{Simon1979,Moiseyev1998}. Our use of perturbative expansions does imply that the field strengths accessible by this effective two-level method are limited, but this is not a problem when compared with recent experiments \cite{Nandi2022}. 

The effective Hamiltonian has eigenstates which are dressed by the field $\ket{\pm}$ , with associated eigenvalues 
\begin{equation}
    \lambda_{\pm} = \frac{h_{aa}+h_{bb}}{2}\pm \frac{1}{2}\sqrt{(h_{aa}-h_{bb})^2+4h_{ab}^2}.
    \label{eq:lambdapm}
\end{equation}
Since $H_{\textrm{eff}}$ is non-Hermitian, its eigenstates will not in general be orthogonal in the standard inner product, i.e. $\innProd{\pm}{\mp}$ is not necessarily zero. The {\it complex} generalized Rabi frequency is defined as: 
\begin{align}
W &= \sqrt{(h_{bb}-h_{aa})^2+4h_{ab}^2}.
\label{eq:W}
\end{align}

When viewed from a semi-classical perspective, the model has the disadvantage that it is restricted to an atom interacting with a {\it flat-top} envelope FEL pulse, which implies that more general features that depend on the FEL envelope are intractable. In future works, however, parameters provided here could be used in more general semi-classical models to study the role of envelope effects, or with more elaborate description of the quantum field.

\subsection{Resolvent operator formalism and subspace considerations}\label{sec:resolvent} 
Consider the evolution of a coherent system governed by the time-independent total Hamiltonian, $H$ in Eq.~(\ref{eq:H_QO}), with the initial condition, $\ket{\Psi(0)} =\ket{a}$ at $t=0$. The time-dependent amplitude for state $\ket{c}$, at $t>0$, is obtained as
\begin{equation}
    U_{ca}(t) = -\frac{1}{2\pi i}\lim_{\eta \to 0^+}\int_{-\infty}^{\infty} dE e^{-iEt} G_{ca}(E+i\eta), 
    \label{eq:U}
\end{equation}
where 
$G_{ca}(z)=\matrixelement{c}{G(z)}{a}$ 
is a matrix element of the resolvent of the total Hamiltonian 
\begin{equation}
G(z)=\frac{1}{(z-H)}, 
\label{eq:Gofz}
\end{equation}
where $\ket{a}$ and $\ket{c}$ are eigenstates of the uncoupled Hamiltonian, $H_0$ \cite{CT1998}. 
%
Subspace projection operators for $\mathcal{P}$ and $\mathcal{Q}$ are defined as 
\begin{equation}
    P = \ket{a}\bra{a} + \ket{b}\bra{b}, \quad Q = \mathbb{1}-P,
\end{equation}
and they satisfy the usual properties of orthogonal projection operators
\begin{equation}
    P^2 = P, \quad Q^2=Q,\quad PQ=QP=0.
\end{equation}
The interaction, $V$, is responsible for the coupling between $\mathcal{P}$ and $\mathcal{Q}$ as  
\begin{equation}
PHQ = PVQ,\,\,\, PH_0Q = QH_0P = 0.
\end{equation}

The model relies on a perturbative expansion of the coupling between the subspaces $\mathcal{P}$ and $\mathcal{Q}$, where both direct coupling   and indirect coupling between between $\ket{a}$ and $\ket{b}$ is treated nonperturbatively \cite{CT1998}. The resolvent equation for subspace $\mathcal{P}$ is 
\begin{equation}
        PG(z)P = P\frac{1}{z-PH_0P - PR(z)P}P,
    \label{eq:resolventP}
\end{equation}
where the {\it level-shift} operator is defined as 
\begin{equation}
    R(z) = V + VQ\frac{1}{z-QH_0Q-QVQ}QV.
\label{eq:Rofz}
\end{equation}
The operator in the denominator of Eq.~(\ref{eq:resolventP}), 
\begin{equation}
   H_\mathrm{eff}= PH_0P + PR(z)P,
   \label{eq:HeffDef}
\end{equation}
is an {\it effective} Hamiltonian for dynamics in subspace $\mathcal{P}$,   
where $PR(z)P$ accounts for effects due to coupling to subspace $\mathcal{Q}$. The matrix elements of the effective Hamiltonian can be written as 
\begin{align}
    \begin{split}
        &h_{aa} = E_a + R_{aa}, \quad h_{bb} = E_b + R_{bb}\\
        &h_{ba} = R_{ba}, \quad h_{ab} = R_{ab},
    \end{split}
\end{align}
where $R_{ij}$ are matrix elements of $R(z).$ We will use the pole-approximation for the matrix elements of $R$, which amounts evaluating them at a fixed energy. This is a good approximation near the resonance \cite{Beers1975,knight_saturation_1977}.

Expansion of the level-shift operator, $R(z)$, in the operator $QVQ$ allows for Eq.~(\ref{eq:Rofz}) to be rewritten as a series   
\begin{align}
\begin{split}
     R(z) &=  V + VQ\frac{1}{z-QH_0Q}QV +\\
    &+ VQ\frac{1}{z-QH_0Q}QVQ\frac{1}{z-QH_0Q}QV + ...,
\end{split}
\label{eq:RofzSeries}
\end{align}
which is diagrammatically exemplified in Fig.~\ref{fig:levels}~(b) up to fourth order. The first diagram (term) corresponds to the usual Rabi interaction in the two-level system: $V_{ba}=V_{ab}=\Omega_0/2$. The second diagram (term) corresponds to second-order AC Stark shift via states outside the two-level system. The second-order term also accounts for depletion of the excited atomic state by photoionization. The third diagram (term) corresponds to a complex correction to the Rabi frequency due to coupling of $\ket{a}$ and $\ket{b}$ via other states. The fourth diagram (term) corresponds to fourth-order AC Stark shift by other states, which is required for a depletion from the ground state due to non-resonant photoionization. All permutations of time-orders of the interactions in the diagrams are included in our calculations (not shown), which implies that transitions  beyond the RWA are taken into account to fourth order in the interaction, or to second order in the intensity of the field: $V^4\propto {\cal E}_0^4\propto I^2$.  

Using Eqs.~(\ref{eq:HeffDef}) and (\ref{eq:RofzSeries}), the elements of the effective Hamiltonian in Eq.~(\ref{eq:Heff}), can now be evaluated, up to any desired order in the power of the field. Introducing the level-shift scaling factors $\rho^{(n)}_{ij}$  to separate the field dependence from the atomic contribution, we can write the matrix elements of the level-shift operator as
\begin{equation}
    R_{ij}=\sum_{n=1}^\infty \frac{{\cal E}_0^{n}}{2^{n}}\rho^{(n)}_{ij}.
    \label{eq:sumR}
\end{equation}
In this paper we include terms up to $n=4$.

The diagonal elements of $R$ can be split into real and imaginary parts, where the real part is responsible for energy shifts, and the imaginary part is related to the ionization rate, 
\begin{equation}
    R_{ii} = \delta_i -i \frac{\gamma_i}{2}, \,\, i\in\{a,b\}.
\end{equation}
In Fig.~\ref{fig:levels}(c) the associated life-times, $\tau_i = 1/\gamma_i$ of the atomic states are shown and two different domains of photoionization, I and II, are indicated, where the atom is predominately ionized from the excited state and ground state, respectively \cite{Nandi2022}. We note that the Rabi period is much smaller than the life-times of the atomic states, $T=2\pi/\Omega<\tau_i$.  

The real part of the off-diagonal elements gives the effective Rabi frequency, while the imaginary part relates coherent transitions between the two states in $\mathcal{P}$, via intermediate continua in $\mathcal{Q}.$ 
As the effective Hamiltonian is complex symmetric, we denote the off-diagonal elements as 
\begin{equation}
    R_{ba} = R_{ab} = \frac{\Omega + i\beta}{2},
    \label{eq:omegabeta}
\end{equation} 
where $\Omega \neq \Omega_0$ since our perturbative expansion of the level-shift operator is to fourth order.
In Fig.~\ref{fig:levels}(d) the AC-Stark shifts due to other states are compared with the Rabi frequency, $\Omega$, and ponderomotive energy, $U_p$, as a function of FEL intensity. Since the excited state has a positive energy shift, $\delta_b>0$, and the ground state has a negative energy shift, $\delta_a<0$, the photon energy needed to fulfill the resonance condition increases with FEL intensity, $\Delta \epsilon_{ba}+\delta_{b}-\delta_{a}$. The role of $U_p$ is to dress the continuum states, but since this effect is smaller than all other effects, it will be neglected in our model. Physically, it would amount to a small red-shift of the photoelectron distribution in kinetic energy by $-U_p$. In the high intensity range we see the Rabi frequency crosses the energy gap to the closest complement atomic state: $\Omega\approx \Delta \epsilon_{bc}$, which implies that the effective two-level model starts to break down. The Rabi frequency is much smaller than the energy difference in the atomic two-level system: $\Omega < \Delta \epsilon_{ba}$, which implies that interactions are within the ``weak-coupling'' regime, and that a break down of Rabi oscillations by ``strong coupling'' can not be reached for a two-level model of the helium atom \cite{Autler1955}. Indeed, clear Rabi oscillations are present in both domain I and II, as shown in  Fig.~\ref{fig:levels}~(e) and (f), respectively. In the high-intensity case, shown in Fig.~\ref{fig:levels}~(f), there is noticeable photoionization losses over a single Rabi period of a few percent, in agreement with the two orders of magnitude ratio between the life-time and Rabi period, see domain II in Fig.~\ref{fig:levels}~(c).         

\subsection{Parameters for the effective Hamiltonian}\label{sec:param}

The scaling factors, $\rho_{ij}^{(n)}$ in Eq.~(\ref{eq:sumR}), can be used to determine the parameters for the effective Hamiltonian in Eq.~(\ref{eq:Heff}) at {\it any} intensity, provided that the two-level model itself is valid. In order to estimate the range of validity for the model, we present in Tab. \ref{tab:conditions} two critical field conditions for hydrogen and helium atoms. The first condition: $\gamma_a=\gamma_b$, implies that both states ionize at the same rate independently, which means that strong quantum interference effects and stabilization may be present. The second condition: $\Omega_0=\Delta\epsilon_{bc}$, implies that the Rabi energy shift is on the order of the level spacing to the closest coupled state $\ket{c}$ from the excited state $\ket{b}$, which means that the effective two-level model is not justified \cite{CT1998}. It is found that the condition for equal rates from both states is not within the realm of allowed intensities for the hydrogen atom. This implies that more advanced few-level effective models are required for quantitative studies of stabilization of hydrogen atoms. In contrast, the equal rates condition is reached for the helium atom within the validity of the two-level model, which implies that quantitative studies of quantum interference and stabilization mechanisms can be performed for the helium atom within the analytically tractable two-level model.   

\begin{table}[h]
\caption{Critical fields for Rabi-cycling hydrogen and helium atom. Intensity is obtained from the electric-field amplitude as: $I$[W/cm$^2$]$=3.51\times10^{16}\times {\cal E}_0^2$[au].}\label{tab:conditions}
\begin{ruledtabular}
\begin{tabular}{c|cc}
Atom &  $\gamma_a=\gamma_b$             &   $\Omega_0 = \Delta\epsilon_{bc}$ \\ \hline
H    &  ${\cal E}_0=0.123$\,au          &      ${\cal E}_0=0.0932$\,au          \\ 
He   &  ${\cal E}_0=0.0558$\,au         &     ${\cal E}_0=0.1645$\,au            
\end{tabular}    
\end{ruledtabular}
\end{table}

In Tab. \ref{tab:polarizabilities} we provide $\rho_{ij}^{(n)}$ parameters  for hydrogen and helium atoms. The scaling factors are computed from first principles, using perturbative expansion at the level of CIS, with ECS to enforce outgoing boundary condition. Expected symmetry between the off-diagonal elements is obtained as: $\rho_{ba}^{(3)}=\rho_{ab}^{(3)}$. 

\begin{table*}[ht!]
\caption{The level-shift scaling factors for helium: $1s^2-1s2p$, and hydrogen: $1s-2p$, with linear polarization (L) and circular polarization (C). The number in parentheses indicate the power of ten, and atomic units are used for all quantities.}\label{tab:polarizabilities}
\begin{ruledtabular}
\begin{tabular}{c|ccccc}
$\rho_{ij}^{(n)}$         & Atom/Pol.   & $n=1$      & $n=2$     & $n=3$   & $n=4$                         \\ \hline
$aa$ & H/L          &  $0$       & $-4.299$                       & $0$                            & $ 1452  -233.1 i $            \\ 
$ab$ & H/L          &    $0.7449$        & $0$                            & $ 54.26  +9.322 i $            & $0$                           \\
$ba$ & H/L           &     $0.7449$        & $0$                            & $ 54.26  +9.322 i $            & $0$                           \\ 
$bb$ & H/L           &  $0$       & $11.70 -0.5223i$               & $0$                            & $ 2888  -95.57 i$             \\  \hline

$aa$ & He/L          &  $0$       & $-2.853$                       & $0$                            & $ 155.6  -10.32 i $           \\ 
$ab$ & He/L          &      $0.4040$       & $0$                            & $ 8.533  +0.2326 i $           & $0$                           \\
$ba$ & He/L          &      $0.4040$      & $0$                            & $ 8.533  +0.2326 i $           & $0$                           \\ 
$bb$ & He/L          &  $0$       & $ 2.010 -7.871i(-3)$           & $0$                            & $ 120.0  -0.2154 i$ 
\\  \hline

$aa$ & He/C          &  $0$       & $-2.853$                       & $0$                            & $ 126.6  -15.25 i $           \\ 
$ab$ & He/C          &      $-0.4040$       & $0$                            & $-5.149  -0.3737 i $           & $0$                           \\
$ba$ & He/C          &      $-0.4040$      & $0$                            & $ -5.149  -0.3737 i $           & $0$                           \\ 
$bb$ & He/C          &  $0$       & $ 1.971 -9.159i(-3)$           & $0$                            & $ 85.35  -1.831i(-3)$           \\
\end{tabular}
\end{ruledtabular}
\end{table*}

In the following we will focus on the helium atom, but first we benchmark our methodology with a comparison of parameters for the effective Hamiltonian given in Ref.~\cite{Holt1983}. We find these older parameters for hydrogen are not correct and must be updated, see Appendix \ref{app:comparison}. In order to verify our results, we have performed additional calculations that confirm our numerical values using the extrapolation method~\cite{cormier_extrapolation_1995}.

\subsection{Two-level photoionization rates}\label{sec:rates}

According to Holt at al. the rate of ionization is given by:
\begin{equation}
\frac{dC}{dt}=\gamma_a|a(t)|^2+\gamma_b|b(t)|^2-2\beta \mathrm{Re}[a^*(t)b(t)],
    \label{eq:ionizationrate}
\end{equation}
which depends on the time-dependent state-resolved population of the atom, $|a(t)|^2$ and $|b(t)|^2$, but also on an interference term that is proportional to the $\beta$-coefficient, see Eq.~(\ref{eq:omegabeta}). This non-linear time-dependent photoionization process is shown in Fig.~\ref{fig:levels}~(e) (and (f)) for regime I (and II), where photoionization is strongest when the atom is in its excited (ground) state. Alternatively, ionization can be described in terms of the dressed states $\ket{\pm}$, with the initial condition: $\ket{a}=c_+\ket{+}+c_-\ket{-}$, as
\begin{align}
\begin{split}
   C(t) =1-|c_+|^2e^{-\gamma_+t} - |c_-|^2e^{-\gamma_-t} \\ -2\textrm{Re} [c_+c_-^* e^{-i(\lambda_+-\lambda_-^*)t}\innProd{-}{+}],
\end{split}
\label{eq:ionprob}
\end{align}
where $\gamma_{\pm} = -2\mathrm{Im}[\lambda_{\pm}]$. Eq.~(\ref{eq:ionprob}) corresponds to Eq.\ (13) in Beers and Armstrong~\cite{Beers1975}, and it is clear that if one of $\gamma_{\pm}$ goes to zero the atom will be stabilized in the corresponding dressed state.
Differentiating Eq.~(\ref{eq:ionprob}) with respect to time gives Eq.~(\ref{eq:ionizationrate}) in the dressed-state picture 
\begin{align}
 \begin{split}
    \frac{dC}{dt} &= \gamma_+|c_+|^2e^{-\gamma_+t} + \gamma_-|c_-|^2e^{-\gamma_-t} \\
    &+ 2\mathrm{Re} [i(\lambda_+-\lambda_-^*)c_+c_-^* e^{-i(\lambda_+-\lambda_-^*)t}\innProd{-}{+}],
 \end{split}
\end{align}
where the first two terms correspond to exponential decay at two different rates, while the third term is an interference term that modulates and decays over time. 
%

Given resonant multi-cycle Rabi oscillations: $t \gg T=2\pi/\Omega$, with negligible photoionization losses: $C(t)\ll 1$,  the period-averaged rate is approximately given by:
\begin{align}
\left< \frac{dC}{dt}\right>_{T} \approx \frac{1}{2}(\gamma_a+\gamma_b) \approx \frac{1}{2}(\gamma_+ + \gamma_-), 
\label{eq:fgr}
\end{align}
which implies a period-averaged linear photoionization probability in accordance with Fermi's Golden Rule \cite{Nandi2022}. The $1/2$ factor in Eq.~(\ref{eq:fgr}) can be interpreted as due to the time-averaged populations of the ground state and excited state, or as due to the equally populated dressed states, which are two alternative views to describe Rabi-cycling atoms with $\Delta\omega=0$.   

\subsection{Two-level photoelectron amplitudes}\label{sec:amplitudes}

According to Beers and Armstrong, the interference term, $\beta$, may cause stabilization against photoionization in the long time limit \cite{Beers1975}. However, this stabilization effect is difficult to measure experimentally because the photoionization rate is non-exponential and changes over time \cite{Beers1975,Holt1983}.  
In this work, we propose that photoelectron spectra can be used to study the stabilization mechanism in the alternative {\it energy} domain, which should be experimentally feasible with modern FEL sources \cite{allaria_highly_2012}. To this end, we now derive expressions for photoelectron spectra within the resolvent operator formalism.  
The resolvent equation for coupling from $\mathcal{P}$ to $\mathcal{Q}$ is \cite{CT1998}
\begin{equation}
        QG(z)P = \frac{1}{z-QHQ}QHPG(z)P
    \label{eq:resolventQ}
\end{equation}
Provided we have an expression for $PG(z)P$, we can derive a perturbative expression for $QG(z)P$, which will be required for finding the photoelectron amplitudes, 
\begin{align}
 \begin{split}
         &QG(z)P = \frac{1}{z-QH_0Q}QVPPG(z)P \\
            &+ \frac{1}{z-QH_0Q}QVQ\frac{1}{z-QH_0Q}QVPPG(z)P + ...
 \end{split}
\end{align}
So the resolvent matrix element for finding a photoelectron in state $\ket{\epsilon} \in \mathcal{Q}$ is then to second order in $V$
\begin{align}
\begin{split}
     &G_{\epsilon a}(z) = \bra{\epsilon}\frac{1}{z-QH_0Q}QVPPG(z)P\ket{a}\\
    &+ \bra{\epsilon}\frac{1}{z-QH_0Q}QVQ\frac{1}{z-QH_0Q}QVPPG(z)P\ket{a}.
    \label{eq:ionG}
\end{split}
\end{align}
The interpretation of this expression is that we first propagate within the space $\mathcal{P}$ with the effective Hamiltonian resulting from the interaction with $\mathcal{Q},$ and then we transition to the space $\mathcal{Q}$ through one interaction, where our perturbative expression allows for one more transition within $\mathcal{Q}.$

Using Eq.~(\ref{eq:U}) to translate Eq.~(\ref{eq:ionG}) into a time-dependent amplitude, we find: 
\begin{widetext}
\begin{align}
    U^{(1)}_{\epsilon a\pm}(t) &= \mp i {\cal E}_0 z_{\epsilon b}\frac{ h_{ab}}{W} \exp[-i\frac{t}{2}(E_{\epsilon}+\lambda_\pm)]\frac{\sin[\frac{t}{2}(E_{\epsilon}-\lambda_\pm)]}{(E_{\epsilon}-\lambda_\pm)} ,
    \label{eq:U1}
    \\
      U^{(2)}_{\epsilon a\pm}(t) &= \mp i {\cal E}_0^2 z_{\epsilon \neq b}\frac{(\lambda_\pm-h_{bb})}{2W} \exp[-i\frac{t}{2}(E_{\epsilon}+\lambda_\pm)]\frac{\sin[\frac{t}{2}(E_{\epsilon}-\lambda_\pm)]}{(E_{\epsilon}-\lambda_\pm)},
    \label{eq:U2}
\end{align}
\end{widetext}
for the two dressed-state eigenvalues in Eq.\,(\ref{eq:lambdapm}) and complex generalized Rabi frequency in Eq.~(\ref{eq:W}). The total first- and second-order amplitudes are: 
\begin{align}
U^{(1)}_{\epsilon a}(t) & = U^{(1)}_{\epsilon a+}(t)+U^{(1)}_{\epsilon a-}(t) \\
U^{(2)}_{\epsilon a}(t) & = U^{(2)}_{\epsilon a+}(t)+U^{(2)}_{\epsilon a-}(t), 
\end{align}
respectively. Short-hand notation for energy: $E_{\epsilon} - \lambda_{\pm} = \epsilon-2\omega-\lambda_{\pm}$ and atomic transition elements: 
\begin{align}
    z_{\epsilon b} &= \bra{\psi_\epsilon}z\ket{\psi_b} 
    \label{eq:zb}
    \\
    z_{\epsilon \neq b} &= \sum_{c\neq b}\frac{\bra{\psi_\epsilon} z \ket{\psi_c}\bra{\psi_c}z\ket{\psi_a}}{E_a-E_c},
    \label{eq:znotb}
\end{align}
are used, where $z$ is the z-component of the position operator used for linear polarization.
In the case of circular polarization the following substitution is made: $z\rightarrow 2^{-1/2}(x + i y)$.   

For sufficiently long times, the amplitudes in Eqs.~(\ref{eq:U1}) and (\ref{eq:U2}) will be peaked near the dressed-state energies $\lambda_{\pm}$ given in  Eq.\,(\ref{eq:lambdapm}). In contrast to our earlier theoretical work \cite{Nandi2022}, which was based on time-dependent perturbation theory, these ionization amplitudes incorporate effects of AC-Stark shifts and depletion. The different signs of the amplitudes in Eqs.~(\ref{eq:U1}) and (\ref{eq:U2}) are essential to interpret the photoelectron interference effects in the AT doublet, but also the phases of the elements of the complex effective Hamiltonian enter and must be considered for a complete description of the interference phenomenon.

\section{Results}\label{sec:results}

We now present our results, which were obtained with the model described in Sec.~\ref{sec:theory}. In Secs.~\ref{sec:domains}-\ref{sec:depletion} we work with linearly polarized light, and study how the AT-doublet depends on intensity, detuning, and interaction time. In Sec.~\ref{sec:stabilization} we study stabilization of one of the dressed states, and show how this effect is sensitive to the polarization of the light. The main result of this section is the identification of a signature in the photoelectron spectrum that originates from stabilization.

\subsection{Photoionization domains}\label{sec:domains}

In Fig.~\ref{fig:a_b} we study the contributions of the two bare states $\ket{a},\ket{b}$ to the photoelectron signal, in the regions I and II where $\gamma_a < \gamma_b$ and $\gamma_a > \gamma_b$, respectively. For Fig.~\ref{fig:a_b} (a) and (b) we used the intensity $I = 1\times10^{12}$ W/cm$^2$, which puts it in region I, while (c) and (d) uses $I = 3\times10^{14}$ W/cm$^2$ which corresponds to region II. The interaction time for the different cases are 1 Rabi period (a), 3/2 Rabi period (b), 1/2 Rabi period (c) and 1 Rabi period (d). 
As shown in previous work, the ultra-fast build-up time of an AT doublet is shorter from the ground state than from the excited state \cite{Nandi2022}. This is due to the different interaction times required for the Rabi amplitudes, $a(t)=\cos(\Omega t/2)$ and $b(t)=-i\sin(\Omega t/2)$, to change sign. Full contrast is obtained when the positive and negative parts of the integrated corresponding Rabi amplitude add up to zero.  More precisely, it takes one (two) complete Rabi period(s) for an AT doublet with full contrast to form from the ground (excited) state in the resonant case.

We can now notice some differences and similarities between the two regions. In both regions the photoelectron signal develops an AT-doublet when the interaction time is sufficiently long (b and d), however the number of Rabi periods it takes for the doublet to be noticeable (on resonance $\hbar\omega = 21.69$ eV) is different between the two regions, in agreement with earlier work~\cite{Nandi2022}. 
Here it is also clear that on these timescales the structure of the AT-doublet is fundamentally different in the two regions, with the AT-doublet in region II displaying a clear asymmetry between the upper and lower components of the doublet as the photon energy is changed, while in region I the components have nearly equal strengths for all photon energies. Further, since the generalized Rabi frequency increases with detuning we observe a transition from a single peak to a double peak with increased  detuning in Fig.~\ref{fig:a_b} (a). 
The AT doublet from the ground state is more difficult to distinguish in the detuned case, due to the asymmetry of the ground state contributions. None-the-less, the formation can be observed as the central photoelectron peak, shown in Fig.~\ref{fig:a_b}~(c), is split into two central peaks, shown in Fig.~\ref{fig:a_b}~(d). Beyond the AT doublet structure, we will refer to ``fringes'' found in the photoelectron distributions as external and internal sidebands.

\subsubsection{External sidebands}
\begin{figure*}
\includegraphics[width=0.75\textwidth]{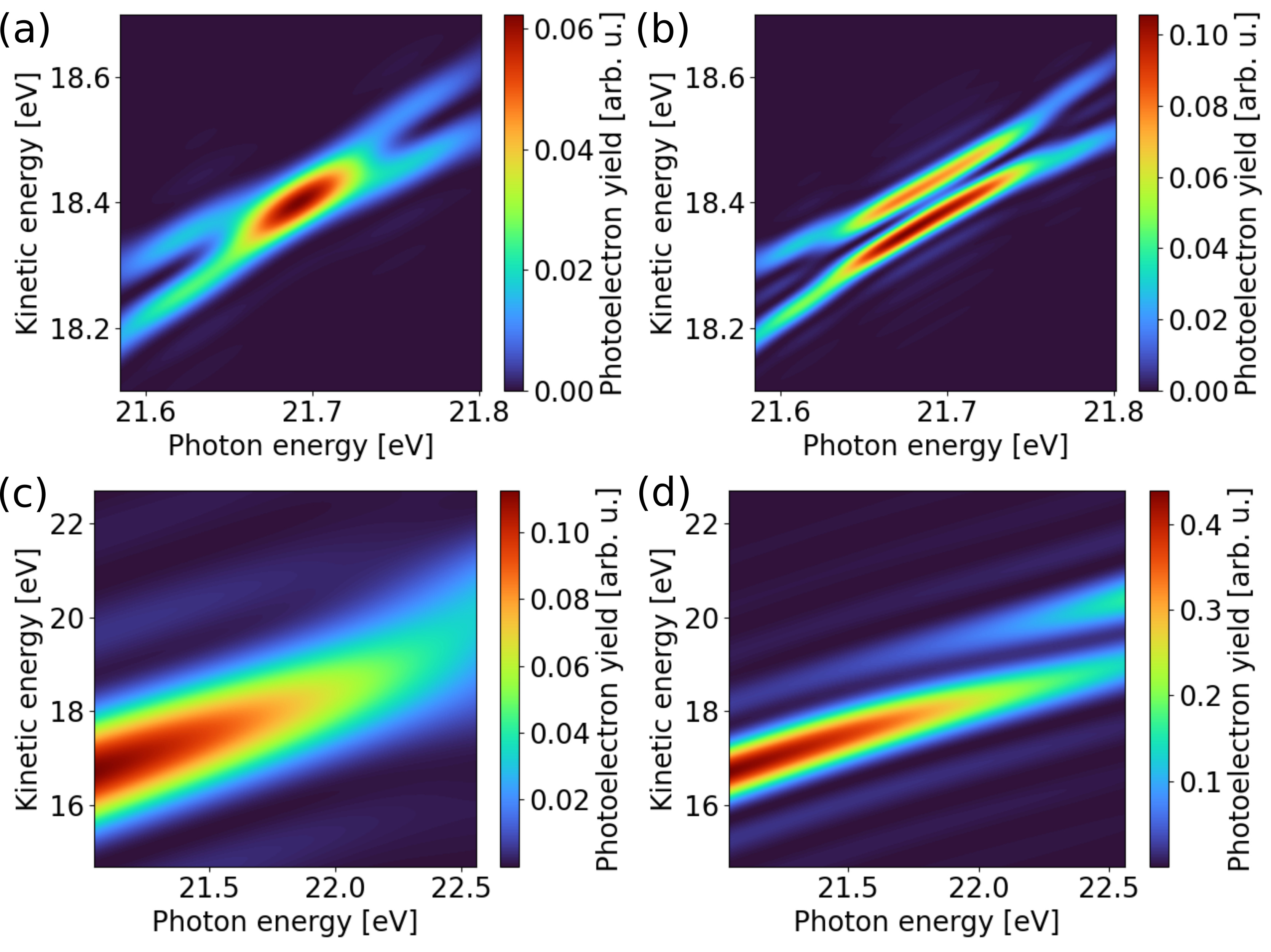}
\caption{\label{fig:a_b} Photoelectron spectrum in the two predicted regions of resonant photoionization. (a) and (b) shows an intensity in region I ($1\times 10^{12}$ W/cm$^2$) for interaction times of 1 and 1.5 Rabi periods, respectively. (c) and (d) shows an intensity in region II ($5\times 10^{14}$ W/cm$^2$) with interaction times of 0.5 and 1 Rabi periods, respectively. The intensity and pulse duration used for each subplot is indicated in Fig.~\ref{fig:levels} (c).}
\end{figure*}

Faint sidebands are observed below and above the main photoelectron structure for all considered cases presented in Fig.~\ref{fig:a_b}. These faint sidebands originate from the oscillatory Fourier transform of the flat-top envelope of the electric field in our quantum model. In the case of a more realistic -- smoother -- envelope function, which could easily be studied within the semi-classical picture, we expect these {\it external} sidebands to be reduced or vanish.   

\subsection{Interference induced asymmetry}\label{sec:interference}

Figure \ref{fig:border} shows the contributions of ionization from the excited state (a), and from the ground state (b) at the intensity where the rates $\gamma_a$ and $\gamma_b$ are equal and the pulse duration corresponds to $N\approx 4$ Rabi periods. In (c) the coherently added signal is shown. Due to interference between the two signals the symmetric AT-doublet in the photoelectron spectrum is shifted away from the resonance condition, towards higher photon energies. 
We stress that it is the relative weight of the probabilities between the upper and lower AT peak that changes, while the position of the AT peaks remains the same. At resonance it is easy to understand this phenomenon using the usual Rabi amplitudes: $a(t) \sim e^{i\Omega t/2}+e^{-i\Omega t/2}$ and $b(t) \sim -e^{i\Omega t/2}+e^{-i\Omega t/2}$, which shows that the ground (excited) state has spectral components with equal (opposite) signs. Thus addition of the contributions will enhance one spectral component and decrease the other (see Supplementary Information of Ref.~\cite{Nandi2022} for more details about photoionization based on the Rabi model). 
As mentioned in Ref.~\cite{Nandi2022}, this shift will in general depend on the exact shape of the FEL-pulse, since an envelope will change the instantaneous ionization rates during the pulse.

\subsubsection{Internal sidebands}
\begin{figure*}
\includegraphics[width=0.9\textwidth]{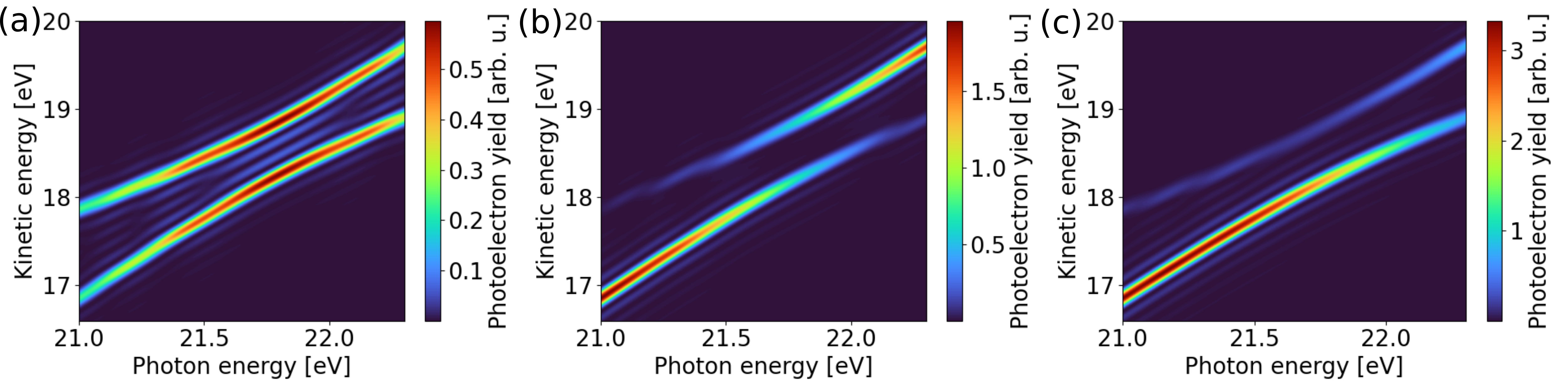}
\caption{\label{fig:border} Photoelectron spectra for an intensity where the ionization rates of the two bare states are of similar magnitude, leading to interference between the resonant and non-resonant ionization paths. (a) shows the contribution from the excited state, (b) the contribution from the ground state, and (c) the combined signal. 
The pulse parameters are $1.09 \times 10^{14}$ W/cm$^2$ and 27.0 fs, which correspond to 4 Rabi cycles in the resonant case. We have also marked this position in parameter space in Fig.~\ref{fig:levels}~(c) with the label ``Fig.~3''.}
\end{figure*}

Similar to our results in Fig.~\ref{fig:a_b}, we observe faint external sidebands in all cases shown in Fig.~\ref{fig:border}. In addition, faint {\it internal} sidebands are observed between the two AT peaks in Fig.~\ref{fig:border}~(a). In contrast to the external sidebands, the manifestation of internal sidebands is a more general phenomenon, which is also present for smooth envelopes \cite{Rogus1986,simonovic_manifestations_2023}. The number of internal sidebands in partial photoionization from the excited state can be interpreted as: $n_b=N-2$, where $N=2,3,...$ is the number of Rabi oscillations performed by the atom. 
In Fig.~\ref{fig:border}~(a) we observe an increase in the number of internal sidebands, from $n_b=2$ to $n_b=4$ ($n_b=3$), for red (blue) detuning, due to the increasing generalized Rabi frequency at the considered pulse parameters. 

We now turn to the question if internal sidebands form from the ground state. 
In Fig.~\ref{fig:border}~(b), we show that the partial photoionization signal from the ground state does {\it not} exhibit internal sidebands at resonance. However, weak asymmetric internal sidebands are found when the laser is detuned. These sidebands look quite similar to the external sidebands. Surprisingly, the total signal, which contains the coherent contributions from both excited and ground state, shows beautiful internal sidebands, as shown in Fig.~\ref{fig:border}~(c). Two internal sidebands are found: $n=2$, which is in agreement with the contribution from the excited state, $n_b=N-2=2$. When the laser is detuned, however, the internal sidebands become asymmetric, reflecting the influence of the photoionization contribution also from the ground state. In contrast to the partial contribution from the excited state, the total photoionization has clear internal sidebands for all detunings (also corresponding to half-integer Rabi periods). This suggests that measurements of internal sidebands could present a way to study rich interference effects between ground and excited states from Rabi cycling atoms.       

\subsection{Depletion induced asymmetry}\label{sec:depletion}

\begin{figure*}
\includegraphics[width=0.85\textwidth]{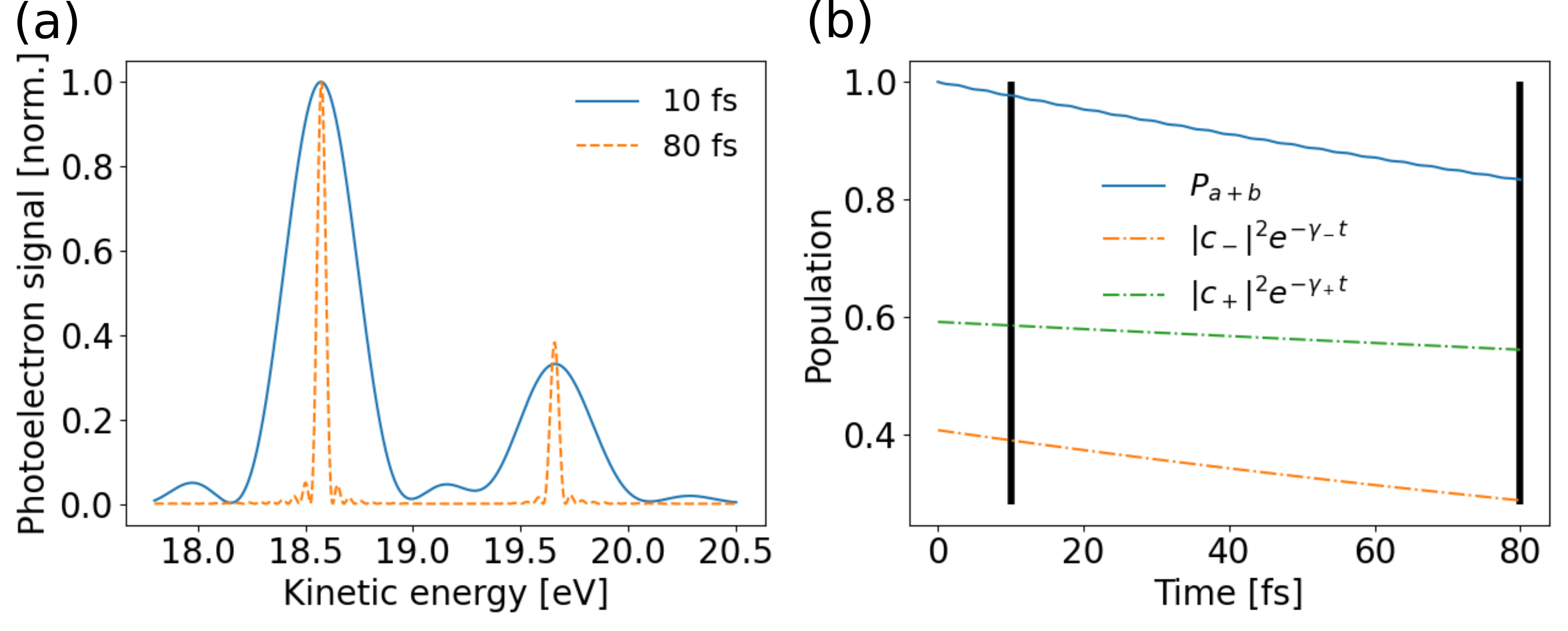}
\caption{\label{fig:asym} (a) Photoelectron distributions for a pulse duration of 10 fs, and a pulse duration of 80 fs. (b) Population left in the two-level system for the 80 fs pulse computed by interaction amplitudes: $P_{a+b}$ and dressed atom amplitudes: $\{|c_+|^2, \, |c_-|^2\}$ weighted by their associated depletion factors. The two vertical lines indicate the pulse durations used in (a).} 
\end{figure*}

Figure~\ref{fig:asym} shows the effect of having different ionization rates for the two dressed states. The strength of the components in the AT-doublet will change over time, if the ionization rate of the two dressed states are different. This is illustrated in Fig.~\ref{fig:asym} (a) for a 10 fs pulse and a 80 fs pulse, at an intensity of $3\times 10^{14}$ W/cm$^2$ and a photon energy of $22.17$ eV. For the 80 fs pulse, the relative strength of the upper component of the AT-doublet has increased when compared to the 10 fs pulse. Initially the component associated to the dressed state with the higher rate will be stronger. However, as  the state with the higher rate becomes depleted, the other component will start to catch up. The initial population of the dressed states will also have an effect on the strength of the components in the doublet. In Fig.~\ref{fig:asym} (b), we can see that the $\ket{+}$ state is initially more populated for this combination of photon energy and intensity. The small oscillations that can be seen in the survival probability come from the cross-term in Eq.~(\ref{eq:ionprob}). Similar oscillations in the ionization probability are seen more clearly in Fig.~\ref{fig:levels} panels (e) and (f).

\subsection{Dressed-atom stabilization}\label{sec:stabilization}

Beers and Armstrong predicted that there should exist combinations of photon energy and intensity for which the ionization rate of one of the dressed states vanishes if both ionization rates in Eq.~(\ref{eq:ionizationrate}) are equal: $\gamma_a=\gamma_b$ \cite{Beers1975}. In their work, as well as in Holt \textit{et al.} \cite{Holt1983}, it was assumed that $\beta = \pm\sqrt{\gamma_a\gamma_b}$. This assumption is valid in the case when there is only a single ionization continuum available, however, it is {\it not} valid when several continua are available, as shown in Appendix \ref{app:beta}. In this work we compute general values of $\beta$, see Eq.~(\ref{eq:omegabeta}), by a systematic expansion of the level-shift operator, see Eq.~(\ref{eq:RofzSeries}). For the systems considered here there are always two continua available for the photoelectron when the FEL pulse is linearly polarized ($s$- and $d$-wave with $m=0$). In contrast, the dipole selection rules for circularly polarized FEL pulses only allow for a single ionization continuum for two-photon ionization ($d$-wave with $|m|= 2$). We find that it is the dressed state $\ket{+}$ that can be stabilized.  
In general it is the relative sign of $\Omega$ and $\beta$ that determines which of the dressed states, $\ket{+}$ or $\ket{-}$, that is stabilized. This point is explained in Appendix \ref{app:beta}.

Figure~\ref{fig:rates} shows a comparison of the ionization rate $\gamma_+$ of $\ket{+}$, as a function of intensity and photon energy for linear and circular polarization. There we can observe that  $\gamma_+$ vanishes for certain FEL parameters for circularly polarized light (along the red line in the plot), which implies that $\ket{+}$ is stabilized. The same phenomena is {\it not} observed for linear polarization in helium. 
%
Consider the photoionization process from the ground $s$-state and excited $p$-state, which for a linearly polarized field lead to final $s$- and $d$-waves. We have showed, in the supplemental information of Ref.~\cite{Nandi2022}, that the relative strength of the resonant and non-resonant ionization processes can be different for $s$- and $d$-waves. Therefore it is not possible to ``rotate'' the $s$- and $d$- continua into a single effective continuum channel for both resonant and non-resonant ionization simultaneously. Thus, linear polarization leads to ``genuine'' multiple continua, with effects beyond the earlier works \cite{Beers1975,Holt1983}.   
This means that stabilization will not happen for linearly polarized light, but it may happen for circular polarization.

Single-continuum stabilization could in principle be observed as a ``dip'' in the ionization probability at a specific intensity and photon energy combination \cite{Beers1975}. However, ionization will still take place from the $\ket{-}$ state, and for short times this will compensate for the lack of ionization from $\ket{+}$, see lower panels in Fig.~\ref{fig:rates} and Eq.~(\ref{eq:fgr}). It is only for pulses long enough to significantly deplete $\ket{-}$ that the stabilization effect can be observed in the total ionization probability. For helium this would require a pulse duration of thousands of Rabi cycles. Since the required intensity for the resonant stabilization is fixed, this implies that the pulse must be 1 000 longer and thus 1 000 times more energetic, which is unrealistic given present technology. Here we present an alternative signature of this stabilization effect, by considering the asymmetry of the AT-doublet in the photolectron spectrum as a function of photon energy at fixed intensity.

\begin{figure*}
\includegraphics[width=\textwidth]{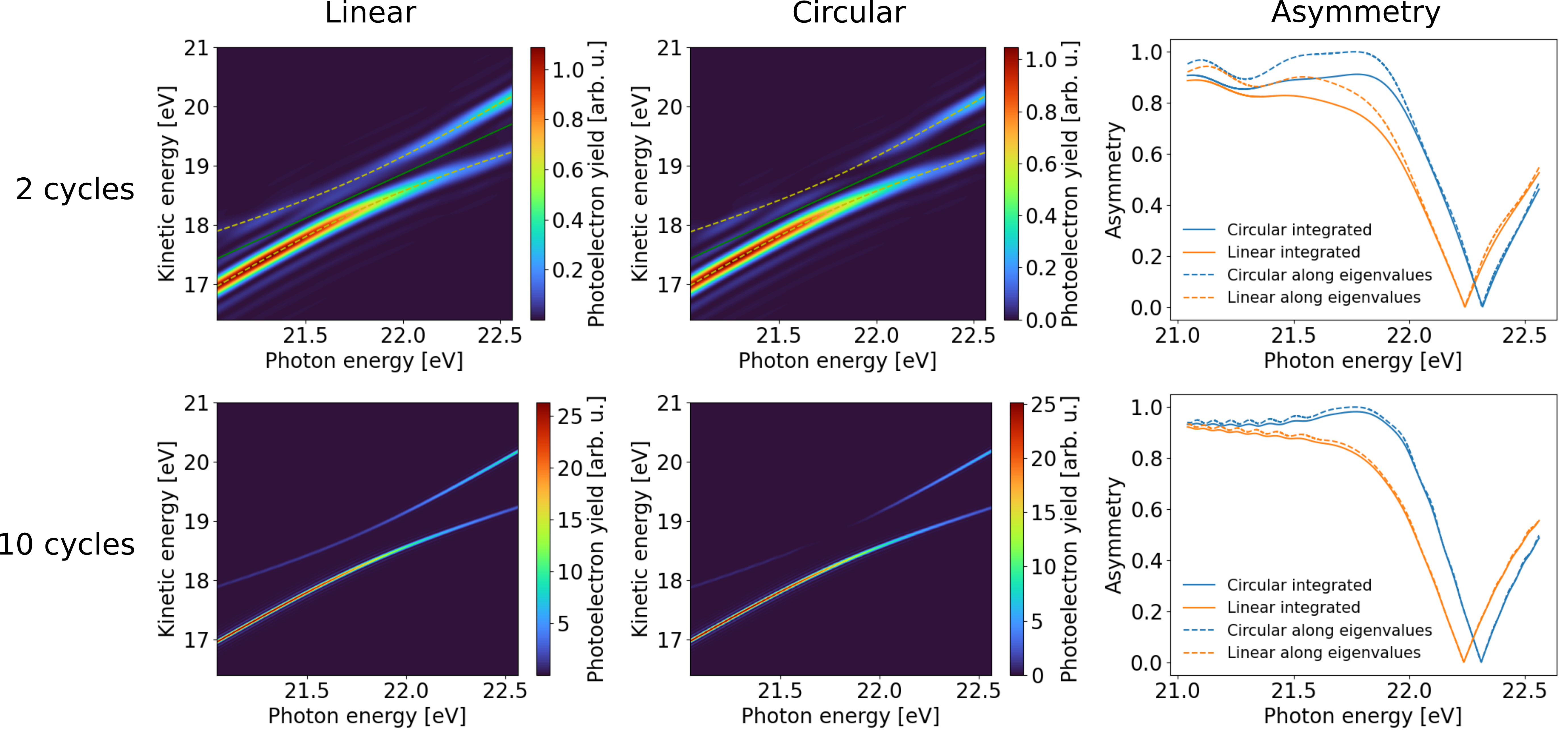}
\caption{\label{fig:circ} Photoelectron spectra for linear (left column) and circular (middle column) polarization at the intensity when the rates $\gamma_a$ and $\gamma_b$ are equal. The interaction time is 2 Rabi cycles (upper row), and 10 Rabi cycles (bottom row), respectively. A shift of the symmetric doublet away from the resonance is observed in both cases. In the case of circular polarization, the higher peak (corresponding to $\lambda_+$) has a ``node'' close to the resonance, which implies stablization of $\ket{+}$. In the case of linear polarization the same effect is not observed. At red detuning, $\omega<21.69\,$eV, the strength of the higher peak is weaker than the lower peak, which implies that photoionization from $\ket{+}$ is lower than from $\ket{-}$. Asymmetry parameters for the doublet is shown in the right column (see main text for details).}
\end{figure*}

In Fig.~\ref{fig:circ} a comparison is made between the photoelectron spectra for  linearly and circularly polarized light, each for the intensity where the ionization rates $\gamma_a$ and $\gamma_b$ are equal. Two pulse durations are considered (2 and 10 Rabi cycles). In the case of circular polarization, a ``node'' in the strength of the upper component of the AT-doublet can be observed close to resonance. This node is associated with lack of ionization from the $\ket{+}$ state. No such node can be seen with linear polarization because $\ket{+}$ is not stabilized for any photon energy.

These phenomena can be studied in more detail by considering the asymmetry of the spectrum with respect to the peaks associated to the eigenvalues $\lambda_{\pm}$. We consider the asymmetry parameter
\begin{equation}
    A = \frac{\left|S_+-S_-\right|}{S_+ + S_-},
\end{equation}
where $S_{\pm}$ is the height of the peak associated to eigenvalue $\lambda_{\pm}$. This would mean measuring the heights along the dashed, yellow lines in Fig.~\ref{fig:circ}. This gives rise to the dashed lines in the right column of Fig.~\ref{fig:circ}. We can see that for a particular photon energy, the asymmetry parameter of the spectrum for circular polarization goes to one, meaning that one of the peaks has vanished. This does not occur for linear polarization. We can also define the $S_{\pm}$ as being the integrated spectrum above or below the  position of the single photoelectron peak (if there was no AT splitting). This is indicated by the green line in the spectra, and the corresponding asymmetry values are indicated by full lines in the asymmetry plots. Due to the presence of strong sidepeaks in the two Rabi cycle case, the asymmetry does not reach one for circular polarization, but a maximum is observed at the same location. As the pulse duration is increased the strength of the sidepeaks is reduced and the integrated definition approaches that of evaluating the peak heights. 

\begin{figure*}
\includegraphics[width=0.9\textwidth]{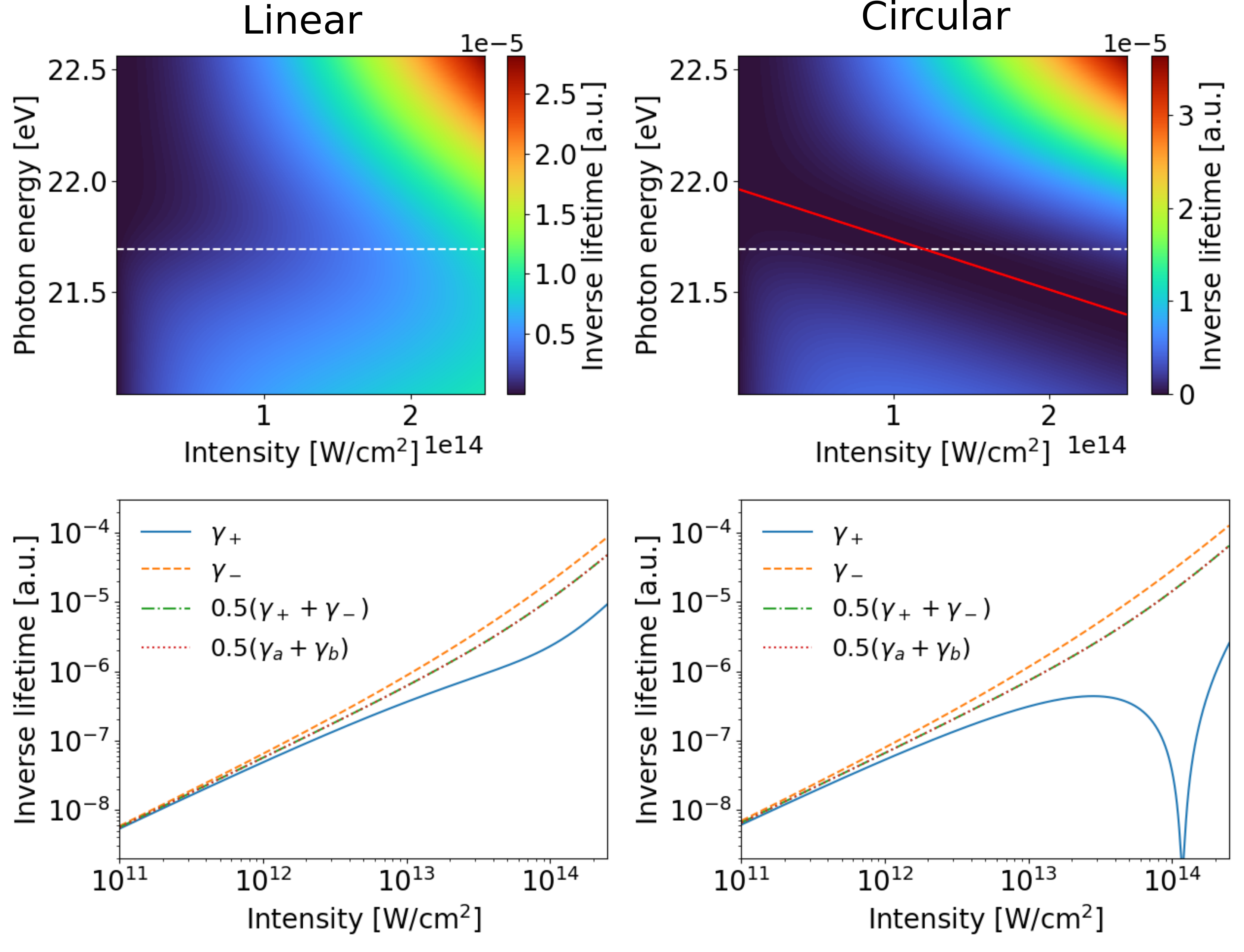}
\caption{\label{fig:rates} Top row: comparison of $\gamma_+$ for linear and circular polarization, as a function of the intensity and the photon energy. The left column corresponds to linear polarization and the right column to circular polarization. The red line in the circular plot shows for which combinations of photon energy and intensity the ionization rate of the $\ket{+}$ state goes to zero. The white dashed line indicates the field-free resonance photon energy. Bottom row: comparison of the ionization rates of the dressed states $\ket{\pm}$ as a function of intensity at the field-free resonance. Also the cycle averaged rates $(\gamma_+ +\gamma_-)/2$ and $(\gamma_a + \gamma_b)/2$ are included for comparison.}
\end{figure*}

\section{Conclusions}\label{sec:conclusions}
In this work we have studied non-perturbative resonant two-photon ionization (1+1) of atoms at XUV-wavelengths using the resolvent operator formalism. Matrix elements of the level-shift operator were calculated using perturbative expansion, up to fourth order in the strength of the electric field, and scaling parameters for helium and hydrogen were presented. It was showed that helium atoms can be stabilized against resonant two-photon ionization with circularly polarized XUV fields, which has a natural explanation in terms of dressed states. Surprisingly, we find that it is the most energetic dressed state, $\ket{+}$, that is stabilized. In contrast, linear polarization does not show this effect, but ionization from one of the dressed states is suppressed relative to the other one.

Our work provides a quantitative physical explanation for {\it why} dressed states may have different ionization rates for linearly polarized XUV pulses, an effect that was first predicted by Girju {\it et al.} using Floquet theory for hydrogen atoms  \cite{Girju2007}. 
The scaling parameter presented in this work allowed us to study a range of non-linear photoionization domains, as originally proposed by Beers and Armstrong \cite{Beers1975}. We computed analytically the photoelectron distributions to the lowest photoelectron peak using the non-Hermitian dynamics of the two-level system as the atomic source. Above-threshold ionization, which has previously been observed by numerical propagation of TDSE \cite{LaGattuta1993,Girju2007}, was not considered in this work, but could be computed by further perturbative expansion in {\cal Q}.

Finally, we have predicted that the stabilization effect can be identified through the asymmetry of the AT-doublet in the photoelectron spectrum, with pulse parameters accessible to seeded FEL sources, such as FERMI~\cite{allaria_highly_2012,Nandi2022}. 
Our most surprising result is that an atom that undergoes ten Rabi cycles exhibits a single photoelectron peak in the stabilization regime. This clearly shows that there is no direct relationship between population of dressed states and the corresponding photoelectron probability distribution in the stabilization regime. 

Partial stabilization of IR dressed autoionizing states has recently been observed in a transient absorption experiment \cite{harkema_autoionizing_2021}. There the stabilization mechanism is attributed to interference between radiative ionization and autoionization processes, whereas the stabilization studied in this work is due to interference within the photoionization process alone. The multiple continua involved in the autoionization process prevents complete stabilization, which similarly to the single peak we observe in the stabilization regime would lead to a single window resonance rather than two resonances with different widths.

We envision that stabilization effects will be interesting to study in future experiments with intense coherent XUV pulses. As an example, Ramsey-like interference structures have been shown by Wollenhaupt {\it et al.} in pump-probe experiments on alkali atoms \cite{Wollenhaupt2003}. Due to phase discontinuities, pump-probe experiments allow for selective preparation of dressed states \cite{Wollenhaupt2006}. In the stabilization regime, we predict that novel interference patterns will be present, compared to previous studies on alkali atoms (in domain I) \cite{Wollenhaupt2003}.   

Although we have restricted our study to monochromatic pulses (a single Fock state in quantum optics), our results are closely related to the control of ionization in two-photon resonant ionization with linearly polarized chirped pulses, first predicted by Saalmann {\it et al.} \cite{Saalmann2018}. This is because Rapid Adiabatic Passage (RAP) triggered by chirped pulses leads to transient population of a dressed state \cite{Wollenhaupt2006}, either $\ket{+}$ or $\ket{-}$, depending on the sign of the chirp of the pulse. Therefore, we predict that RAP with {\it circularly} polarized chirped pulses can be used to further improve control of photoionization, by taking advantage of transient stabilization, which may find future applications in quantum control experiments.

\begin{acknowledgments}
We acknowledge Saikat Nandi, Mattias Bertolino and Ulf Saalmann for useful discussions. JMD acknowledges support from the Swedish Research Council: 2018-03845,
the Olle Engkvist Foundation: 194-0734 and the Knut and Alice Wallenberg
Foundation: 2017.0104 and 2019.0154. 
\end{acknowledgments}

\appendix
\section{Effective Hamiltonian parameters for hydrogen}\label{app:comparison}

Here we study the parameters of the effective Hamiltonian for light resonant with 2p in hydrogen. A comparison is made with our perturbative approach and that of Ref.~\cite{Holt1983}, where the analytically continued non-Hermitian Floquet Hamiltonian was used~\cite{maquet_stark_1983}. The results are presented in Tab.~\ref{tab:comparison}. At a field strength of ${\cal E}_0 = 0.001$ a.u., there is good agreement for all parameters except $\gamma_a$ and $\beta$. As the field strength is increased the agreement is worse for all parameters, but still $\gamma_a$ has the biggest discrepancy. The disagreement with tabulated values could potentially be explained by limited numerical accuracy for especially $\gamma_a$, since it is acknowledged by Holt et al. that their values of $\gamma_a$ and $\beta$ are potentially unreliable \cite{Holt1983}. Comparing our values for the scaling parameters for hydrogen with the values extracted by comparison with TDSE calculations in Ref.~\cite{Zhang2022} gives a fair agreement ($\approx 10-35\%$ relative difference) for $\rho_{aa}^{(2)}$, $\rho_{bb}^{(2)}$, and $\textrm{Im}[\rho_{aa}^{(4)}]$. The third-order off-diagonal contribution, e.g. $\beta$ which is essential for stabilization, and fourth order shifts for $\ket{b}$ are not included in their model. We note that often used ``local approximation'' in time-dependent essential state models, does not either include off-diagonal complex contributions because its original application was to ``weak'' fields \cite{Pahl1996}. In this work we considered ``strong'' fields, which give rise to Rabi oscillations and non-linear ways to photoionize the atom. In this case, complex off-diagonal matrix elements should be used, also in time-dependent essential state models, to describe interference effects to the continuum.  

\begin{table*}[ht!]
\caption{Parameters of effective Hamiltonian for the case of resonant two-photon ionization of H via the 2p state and He via the 1s2p state. The number in parentheses indicate the power of ten, and atomic units are used for all quantities.}\label{tab:comparison}
\begin{ruledtabular}
\begin{tabular}{cccccccc}
 Atom & ${\cal E}_0$ & $\omega $ & $\delta_a$ & $\delta_b$ & $\gamma_a$ & $\gamma_b$ & $\beta$\\ \hline 
H\footnotemark[1] & 0.001 & $0.375$ &$-1.0748(-6)$ & $2.9244(-6)$ & $2.9135(-11)$& $2.6118(-7)$ & $2.3304(-9)$\\
H\footnotemark[2] &       &         & $-1.07(-6)$ & $2.92(-6)$& $6.0(-10)$ & $2.61(-7)$ & $2.9(-9)$ \\
H\footnotemark[1] & 0.005 &         & $-2.6815(-5)$ & $7.3218(-5)$ & $1.8209(-8)$& $6.5368(-6)$ & $2.9130(-7)$\\
H\footnotemark[2] &       &         & $-2.54(-5)$ & $7.17(-5)$& $9.1(-8)$ & $6.54(-6)$ & $3.6(-7)$  \\
H\footnotemark[1] & 0.01  &         & $-1.0658(-4)$ & $2.9423(-4)$ & $2.9135(-7)$& $2.6237(-5)$ & $2.3304(-6)$\\
H\footnotemark[2] &       &         & $-9.59(-5)$ & $2.81(-4)$& $8.8(-7)$ & $2.55(-5)$ & $2.8(-6)$  \\
\hline
He\footnotemark[1] & 0.001 & $0.7972$ & $-7.1329(-7)$ & $5.0258(-7)$& $1.2901(-12)$ & $3.9355(-9)$ & $5.8140(-11)$ \\
He\footnotemark[1] & 0.005 &          & $-1.7826(-5)$ & $1.2569(-5)$& $8.0629(-10)$ & $9.8403(-8)$ & $7.2675(-9)$ \\
He\footnotemark[1] & 0.01  &          & $-7.1233(-5)$ & $5.0332(-5)$& $1.2901(-8)$ & $3.9381(-7)$ & $5.8140(-8)$ \\
\end{tabular}
\end{ruledtabular}
\footnotetext[1]{This work}
\footnotetext[2]{Reference \cite{Holt1983}} \\
\end{table*} 

\section{Role of \texorpdfstring{$\beta$}{beta} in stabilization of dressed states}\label{app:beta}
In this appendix we clarify the role of $\beta$ in the stabilization of dressed states. We start by deriving an expression for $\beta$. The amplitude $M$ for the third-order process in Fig.~\ref{fig:levels} (b), which is responsible for the imaginary part of $R_{ba}$, is given by
\begin{equation}
M = \sum_{c\neq b,c^{\prime}}\frac{\bra{b}V\ket{c^{\prime}}\bra{c^{\prime}}V\ket{c}\bra{c}V\ket{a}}{(E_a + i\eta -E_{c^{\prime}})(E_a + i\eta-E_{c})}. 
\label{eq:pertthird}
\end{equation}
In terms of the atomic states and energies, Eq.~(\ref{eq:pertthird}) can be written as
\begin{equation}
M = \sum_{c\neq b,c^{\prime}}\frac{E_0^3\bra{\psi_b}z\ket{\psi_{c^{\prime}}}\bra{\psi_{c^{\prime}}}z\ket{\psi_c}\bra{\psi_c}z\ket{\psi_a}}{2^3(\epsilon_a + 2\omega + i\eta -\epsilon_{c^{\prime}})(\epsilon_a + \omega + i\eta-\epsilon_{c})}. 
\label{eq:pertthirdatom}
\end{equation}
Here linear polarization has been assumed, and in the case of circular polarization the $z$ operator should be replaced by $2^{-1/2}(x+iy)$ for absorption and $2^{-1/2}(x-iy)$ for emission. When taking the $\eta \to 0^+$ limit, the denominator involving $\epsilon_{c^{\prime}}$ will lead to a pole in the continuum and Eq.~(\ref{eq:pertthirdatom}) becomes
\begin{align}
\begin{split}
      M &=   \textrm{p.v.} \frac{E_0^3}{2^3} \sum_{c\neq b, c^{\prime}} \frac{\bra{\psi_b}z\ket{\psi_{c^{\prime}}}\bra{\psi_{c^{\prime}}}z\ket{\psi_c}\bra{\psi_c}z\ket{\psi_a}}{(\epsilon_a + 2\omega -\epsilon_{c^{\prime}})(\epsilon_a + \omega -\epsilon_{c})}\\ &-i\pi\frac{E_0^3}{2^3}\sum_{c\neq b}\sum_{\ell} \frac{\bra{\psi_b}z\ket{\psi_{\epsilon}^{\ell}}\bra{\psi_{\epsilon}^{\ell}}z\ket{\psi_c}\bra{\psi_c}z\ket{\psi_a}}{\epsilon_a + \omega -\epsilon_{c}},
\end{split}
\end{align}
where the sum over $\ell$ corresponds to a sum over the available continua and p.v.\ denotes the principal value. Following Eqs.~(\ref{eq:zb}) and (\ref{eq:znotb}), the imaginary part of $M$ can be expressed as
\begin{equation}
    \text{Im}[M] = -\pi\frac{E_0^3}{2^3}\sum_{\ell} z_{b\epsilon}^{\ell}z_{\epsilon\neq b}^{\ell}.
\end{equation}
Since $\beta = 2\text{Im}[M]$, we can write
\begin{equation}
    \beta = -\pi\frac{E_0^3}{2^2}\sum_{\ell} z_{b\epsilon}^{\ell}z_{\epsilon\neq b}^{\ell} = -\sum_{\ell} \sigma_{\ell}\sqrt{\gamma_b^{\ell}\gamma_a^{\ell}},
\end{equation}
where $\gamma_i^{\ell}$ is the partial rate for ionization from state $i\in\{a,b\}$ to continuum $\ell$, and $\sigma_{\ell}$ is the sign of $z_{b\epsilon}^{\ell} z_{\epsilon\neq b}^{\ell}$. When only one ionization continuum is available $\beta = \pm \sqrt{\gamma_a\gamma_b}$, but this does not hold in general.

Next, we consider the dressed-state energies, $\lambda_{\pm}$, given in Eq.~(\ref{eq:lambdapm}). 
For simplicity we consider the case $h_{aa} = h_{bb}$ which means no detuning, 
\begin{align}
E_a+\delta_a = E_b+\delta_b = E,
\end{align}
and that the ionization rates are equal, 
\begin{align}
\gamma_b = \sum_{\ell}\gamma_b^{\ell} = \gamma_a = \sum_{\ell}\gamma_a^{\ell}  = \gamma.
\end{align}
The dressed-state energies are then reduced to 
\begin{equation}
    \lambda_{\pm} = E -i\frac{\gamma}{2} \pm \frac{1}{2}\sqrt{(\Omega + i\beta)^2}.
\end{equation}
The square root can be written (using the principal branch) as
\begin{equation}
    \sqrt{(\Omega + i\beta)^2} = |\Omega| +\textrm{sgn}(\Omega) i \beta, 
\end{equation}
so that $\lambda_{\pm}$ takes the form
\begin{equation}
    \lambda_{\pm} = E-i\frac{\gamma}{2} \pm \frac{1}{2}\left\{|\Omega| +\textrm{sgn}(\Omega)i\beta \right\}.
\end{equation}
Stabilization occurs when the imaginary part of one of the dressed energies vanishes, i.e. if
\begin{equation}
    \gamma = \pm \textrm{sgn}(\Omega)\beta.
    \label{eq:stabilization}
\end{equation}
Since $\gamma$ is a positive number, it follows that stabilization can occur in only one of the dressed states. If $\Omega$ and $\beta$ have the same sign, then the $\ket{+}$ state with $\lambda_+$ will be stabilized, otherwise the $\ket{-}$ with $\lambda_-$ has the possibility to be stabilized. In the circular polarization case considered in the main text $\Omega$ and $\beta$ are both negative, which explains why the stabilization occurs in the more energetic $\ket{+}$ state, see Fig.~\ref{fig:rates} (right column). For Eq.~(\ref{eq:stabilization}) to hold in the case of multiple continua we must have $\gamma_a^{\ell} = \gamma_b^{\ell}$ for every $\ell$, and $\sigma_{\ell} = \sigma_{\ell^{\prime}}$ for all $\ell$ and $\ell^{\prime}$.

In the case of linear polarization, both hydrogen and helium have positive values for $\Omega$ and $\beta$, which implies that it is again the $\ket{+}$ state that could be stabilized. However, two different continua are present, and Eq.~(\ref{eq:stabilization}) is not satisfied. This results in a reduction of $\gamma_+$, when compared with $\gamma_-$, but it does not go to zero, see Fig.~\ref{fig:rates} (left column).  

\bibliography{bibliography} 

\begin{thebibliography}{43}%
\makeatletter
\providecommand \@ifxundefined [1]{%
 \@ifx{#1\undefined}
}%
\providecommand \@ifnum [1]{%
 \ifnum #1\expandafter \@firstoftwo
 \else \expandafter \@secondoftwo
 \fi
}%
\providecommand \@ifx [1]{%
 \ifx #1\expandafter \@firstoftwo
 \else \expandafter \@secondoftwo
 \fi
}%
\providecommand \natexlab [1]{#1}%
\providecommand \enquote  [1]{``#1''}%
\providecommand \bibnamefont  [1]{#1}%
\providecommand \bibfnamefont [1]{#1}%
\providecommand \citenamefont [1]{#1}%
\providecommand \href@noop [0]{\@secondoftwo}%
\providecommand \href [0]{\begingroup \@sanitize@url \@href}%
\providecommand \@href[1]{\@@startlink{#1}\@@href}%
\providecommand \@@href[1]{\endgroup#1\@@endlink}%
\providecommand \@sanitize@url [0]{\catcode `\\12\catcode `\$12\catcode
  `\&12\catcode `\#12\catcode `\^12\catcode `\_12\catcode `\%12\relax}%
\providecommand \@@startlink[1]{}%
\providecommand \@@endlink[0]{}%
\providecommand \url  [0]{\begingroup\@sanitize@url \@url }%
\providecommand \@url [1]{\endgroup\@href {#1}{\urlprefix }}%
\providecommand \urlprefix  [0]{URL }%
\providecommand \Eprint [0]{\href }%
\providecommand \doibase [0]{https://doi.org/}%
\providecommand \selectlanguage [0]{\@gobble}%
\providecommand \bibinfo  [0]{\@secondoftwo}%
\providecommand \bibfield  [0]{\@secondoftwo}%
\providecommand \translation [1]{[#1]}%
\providecommand \BibitemOpen [0]{}%
\providecommand \bibitemStop [0]{}%
\providecommand \bibitemNoStop [0]{.\EOS\space}%
\providecommand \EOS [0]{\spacefactor3000\relax}%
\providecommand \BibitemShut  [1]{\csname bibitem#1\endcsname}%
\let\auto@bib@innerbib\@empty
\bibitem [{\citenamefont {Huang}\ \emph {et~al.}(2021)\citenamefont {Huang},
  \citenamefont {Deng}, \citenamefont {Liu}, \citenamefont {Wang},\ and\
  \citenamefont {Zhao}}]{huang_features_2021}%
  \BibitemOpen
  \bibfield  {author} {\bibinfo {author} {\bibfnamefont {N.}~\bibnamefont
  {Huang}}, \bibinfo {author} {\bibfnamefont {H.}~\bibnamefont {Deng}},
  \bibinfo {author} {\bibfnamefont {B.}~\bibnamefont {Liu}}, \bibinfo {author}
  {\bibfnamefont {D.}~\bibnamefont {Wang}},\ and\ \bibinfo {author}
  {\bibfnamefont {Z.}~\bibnamefont {Zhao}},\ }\bibfield  {title} {\bibinfo
  {title} {Features and futures of {X}-ray free-electron lasers},\ }\href
  {https://doi.org/10.1016/j.xinn.2021.100097} {\bibfield  {journal} {\bibinfo
  {journal} {The Innovation}\ }\textbf {\bibinfo {volume} {2}},\ \bibinfo
  {pages} {100097} (\bibinfo {year} {2021})}\BibitemShut {NoStop}%
\bibitem [{\citenamefont {Ackermann}\ \emph {et~al.}(2007)\citenamefont
  {Ackermann}, \citenamefont {Asova}, \citenamefont {Ayvazyan}, \citenamefont
  {Azima}, \citenamefont {Baboi}, \citenamefont {B\"{a}hr}, \citenamefont
  {Balandin}, \citenamefont {Beutner}, \citenamefont {Brandt}, \citenamefont
  {Bolzmann} \emph {et~al.}}]{ackermann_operation_2007}%
  \BibitemOpen
  \bibfield  {author} {\bibinfo {author} {\bibfnamefont {W.}~\bibnamefont
  {Ackermann}}, \bibinfo {author} {\bibfnamefont {G.}~\bibnamefont {Asova}},
  \bibinfo {author} {\bibfnamefont {V.}~\bibnamefont {Ayvazyan}}, \bibinfo
  {author} {\bibfnamefont {A.}~\bibnamefont {Azima}}, \bibinfo {author}
  {\bibfnamefont {N.}~\bibnamefont {Baboi}}, \bibinfo {author} {\bibfnamefont
  {J.}~\bibnamefont {B\"{a}hr}}, \bibinfo {author} {\bibfnamefont
  {V.}~\bibnamefont {Balandin}}, \bibinfo {author} {\bibfnamefont
  {B.}~\bibnamefont {Beutner}}, \bibinfo {author} {\bibfnamefont
  {A.}~\bibnamefont {Brandt}}, \bibinfo {author} {\bibfnamefont
  {A.}~\bibnamefont {Bolzmann}}, \emph {et~al.},\ }\bibfield  {title} {\bibinfo
  {title} {Operation of a free-electron laser from the extreme ultraviolet to
  the water window},\ }\href {https://doi.org/10.1038/nphoton.2007.76}
  {\bibfield  {journal} {\bibinfo  {journal} {Nature Photonics}\ }\textbf
  {\bibinfo {volume} {1}},\ \bibinfo {pages} {336} (\bibinfo {year}
  {2007})}\BibitemShut {NoStop}%
\bibitem [{\citenamefont {Emma}\ \emph {et~al.}(2010)\citenamefont {Emma},
  \citenamefont {Akre}, \citenamefont {Arthur}, \citenamefont {Bionta},
  \citenamefont {Bostedt}, \citenamefont {Bozek}, \citenamefont {Brachmann},
  \citenamefont {Bucksbaum}, \citenamefont {Coffee}, \citenamefont {Decker}
  \emph {et~al.}}]{emma_first_2010}%
  \BibitemOpen
  \bibfield  {author} {\bibinfo {author} {\bibfnamefont {P.}~\bibnamefont
  {Emma}}, \bibinfo {author} {\bibfnamefont {R.}~\bibnamefont {Akre}}, \bibinfo
  {author} {\bibfnamefont {J.}~\bibnamefont {Arthur}}, \bibinfo {author}
  {\bibfnamefont {R.}~\bibnamefont {Bionta}}, \bibinfo {author} {\bibfnamefont
  {C.}~\bibnamefont {Bostedt}}, \bibinfo {author} {\bibfnamefont
  {J.}~\bibnamefont {Bozek}}, \bibinfo {author} {\bibfnamefont
  {A.}~\bibnamefont {Brachmann}}, \bibinfo {author} {\bibfnamefont
  {P.}~\bibnamefont {Bucksbaum}}, \bibinfo {author} {\bibfnamefont
  {R.}~\bibnamefont {Coffee}}, \bibinfo {author} {\bibfnamefont {F.-J.}\
  \bibnamefont {Decker}}, \emph {et~al.},\ }\bibfield  {title} {\bibinfo
  {title} {First lasing and operation of an ångstrom-wavelength free-electron
  laser},\ }\href {https://doi.org/10.1038/nphoton.2010.176} {\bibfield
  {journal} {\bibinfo  {journal} {Nature Photon}\ }\textbf {\bibinfo {volume}
  {4}},\ \bibinfo {pages} {641} (\bibinfo {year} {2010})}\BibitemShut {NoStop}%
\bibitem [{\citenamefont {Ishikawa}\ \emph {et~al.}(2012)\citenamefont
  {Ishikawa}, \citenamefont {Aoyagi}, \citenamefont {Asaka}, \citenamefont
  {Asano}, \citenamefont {Azumi}, \citenamefont {Bizen}, \citenamefont {Ego},
  \citenamefont {Fukami}, \citenamefont {Fukui}, \citenamefont {Furukawa} \emph
  {et~al.}}]{ishikawa_compact_2012}%
  \BibitemOpen
  \bibfield  {author} {\bibinfo {author} {\bibfnamefont {T.}~\bibnamefont
  {Ishikawa}}, \bibinfo {author} {\bibfnamefont {H.}~\bibnamefont {Aoyagi}},
  \bibinfo {author} {\bibfnamefont {T.}~\bibnamefont {Asaka}}, \bibinfo
  {author} {\bibfnamefont {Y.}~\bibnamefont {Asano}}, \bibinfo {author}
  {\bibfnamefont {N.}~\bibnamefont {Azumi}}, \bibinfo {author} {\bibfnamefont
  {T.}~\bibnamefont {Bizen}}, \bibinfo {author} {\bibfnamefont
  {H.}~\bibnamefont {Ego}}, \bibinfo {author} {\bibfnamefont {K.}~\bibnamefont
  {Fukami}}, \bibinfo {author} {\bibfnamefont {T.}~\bibnamefont {Fukui}},
  \bibinfo {author} {\bibfnamefont {Y.}~\bibnamefont {Furukawa}}, \emph
  {et~al.},\ }\bibfield  {title} {\bibinfo {title} {A compact {X}-ray
  free-electron laser emitting in the sub-{\aa}ngstr\"{o}m region},\ }\href
  {https://doi.org/10.1038/nphoton.2012.141} {\bibfield  {journal} {\bibinfo
  {journal} {Nature Photon}\ }\textbf {\bibinfo {volume} {6}},\ \bibinfo
  {pages} {540} (\bibinfo {year} {2012})}\BibitemShut {NoStop}%
\bibitem [{\citenamefont {Allaria}\ \emph {et~al.}(2012)\citenamefont
  {Allaria}, \citenamefont {Appio}, \citenamefont {Badano}, \citenamefont
  {Barletta}, \citenamefont {Bassanese}, \citenamefont {Biedron}, \citenamefont
  {Borga}, \citenamefont {Busetto}, \citenamefont {Castronovo}, \citenamefont
  {Cinquegrana} \emph {et~al.}}]{allaria_highly_2012}%
  \BibitemOpen
  \bibfield  {author} {\bibinfo {author} {\bibfnamefont {E.}~\bibnamefont
  {Allaria}}, \bibinfo {author} {\bibfnamefont {R.}~\bibnamefont {Appio}},
  \bibinfo {author} {\bibfnamefont {L.}~\bibnamefont {Badano}}, \bibinfo
  {author} {\bibfnamefont {W.~A.}\ \bibnamefont {Barletta}}, \bibinfo {author}
  {\bibfnamefont {S.}~\bibnamefont {Bassanese}}, \bibinfo {author}
  {\bibfnamefont {S.~G.}\ \bibnamefont {Biedron}}, \bibinfo {author}
  {\bibfnamefont {A.}~\bibnamefont {Borga}}, \bibinfo {author} {\bibfnamefont
  {E.}~\bibnamefont {Busetto}}, \bibinfo {author} {\bibfnamefont
  {D.}~\bibnamefont {Castronovo}}, \bibinfo {author} {\bibfnamefont
  {P.}~\bibnamefont {Cinquegrana}}, \emph {et~al.},\ }\bibfield  {title}
  {\bibinfo {title} {Highly coherent and stable pulses from the {FERMI} seeded
  free-electron laser in the extreme ultraviolet},\ }\href
  {https://doi.org/10.1038/nphoton.2012.233} {\bibfield  {journal} {\bibinfo
  {journal} {Nature Photon}\ }\textbf {\bibinfo {volume} {6}},\ \bibinfo
  {pages} {699} (\bibinfo {year} {2012})}\BibitemShut {NoStop}%
\bibitem [{\citenamefont {Mirian}\ \emph {et~al.}(2021)\citenamefont {Mirian},
  \citenamefont {Di~Fraia}, \citenamefont {Spampinati}, \citenamefont
  {Sottocorona}, \citenamefont {Allaria}, \citenamefont {Badano}, \citenamefont
  {Danailov}, \citenamefont {Demidovich}, \citenamefont {De~Ninno},
  \citenamefont {Di~Mitri} \emph {et~al.}}]{mirian_generation_2021}%
  \BibitemOpen
  \bibfield  {author} {\bibinfo {author} {\bibfnamefont {N.~S.}\ \bibnamefont
  {Mirian}}, \bibinfo {author} {\bibfnamefont {M.}~\bibnamefont {Di~Fraia}},
  \bibinfo {author} {\bibfnamefont {S.}~\bibnamefont {Spampinati}}, \bibinfo
  {author} {\bibfnamefont {F.}~\bibnamefont {Sottocorona}}, \bibinfo {author}
  {\bibfnamefont {E.}~\bibnamefont {Allaria}}, \bibinfo {author} {\bibfnamefont
  {L.}~\bibnamefont {Badano}}, \bibinfo {author} {\bibfnamefont {M.~B.}\
  \bibnamefont {Danailov}}, \bibinfo {author} {\bibfnamefont {A.}~\bibnamefont
  {Demidovich}}, \bibinfo {author} {\bibfnamefont {G.}~\bibnamefont
  {De~Ninno}}, \bibinfo {author} {\bibfnamefont {S.}~\bibnamefont {Di~Mitri}},
  \emph {et~al.},\ }\bibfield  {title} {\bibinfo {title} {Generation and
  measurement of intense few-femtosecond superradiant extreme-ultraviolet
  free-electron laser pulses},\ }\href
  {https://doi.org/10.1038/s41566-021-00815-w} {\bibfield  {journal} {\bibinfo
  {journal} {Nat. Photon.}\ }\textbf {\bibinfo {volume} {15}},\ \bibinfo
  {pages} {523} (\bibinfo {year} {2021})}\BibitemShut {NoStop}%
\bibitem [{\citenamefont {Young}\ \emph {et~al.}(2010)\citenamefont {Young},
  \citenamefont {Kanter}, \citenamefont {Kr\"{a}ssig}, \citenamefont {Li},
  \citenamefont {March}, \citenamefont {Pratt}, \citenamefont {Santra},
  \citenamefont {Southworth}, \citenamefont {Rohringer}, \citenamefont
  {DiMauro} \emph {et~al.}}]{young_femtosecond_2010}%
  \BibitemOpen
  \bibfield  {author} {\bibinfo {author} {\bibfnamefont {L.}~\bibnamefont
  {Young}}, \bibinfo {author} {\bibfnamefont {E.~P.}\ \bibnamefont {Kanter}},
  \bibinfo {author} {\bibfnamefont {B.}~\bibnamefont {Kr\"{a}ssig}}, \bibinfo
  {author} {\bibfnamefont {Y.}~\bibnamefont {Li}}, \bibinfo {author}
  {\bibfnamefont {A.~M.}\ \bibnamefont {March}}, \bibinfo {author}
  {\bibfnamefont {S.~T.}\ \bibnamefont {Pratt}}, \bibinfo {author}
  {\bibfnamefont {R.}~\bibnamefont {Santra}}, \bibinfo {author} {\bibfnamefont
  {S.~H.}\ \bibnamefont {Southworth}}, \bibinfo {author} {\bibfnamefont
  {N.}~\bibnamefont {Rohringer}}, \bibinfo {author} {\bibfnamefont {L.~F.}\
  \bibnamefont {DiMauro}}, \emph {et~al.},\ }\bibfield  {title} {\bibinfo
  {title} {Femtosecond electronic response of atoms to ultra-intense
  {X}-rays},\ }\href {https://doi.org/10.1038/nature09177} {\bibfield
  {journal} {\bibinfo  {journal} {Nature}\ }\textbf {\bibinfo {volume} {466}},\
  \bibinfo {pages} {56} (\bibinfo {year} {2010})}\BibitemShut {NoStop}%
\bibitem [{\citenamefont {Rudenko}\ \emph {et~al.}(2017)\citenamefont
  {Rudenko}, \citenamefont {Inhester}, \citenamefont {Hanasaki}, \citenamefont
  {Li}, \citenamefont {Robatjazi}, \citenamefont {Erk}, \citenamefont {Boll},
  \citenamefont {Toyota}, \citenamefont {Hao}, \citenamefont {Vendrell} \emph
  {et~al.}}]{rudenko_femtosecond_2017}%
  \BibitemOpen
  \bibfield  {author} {\bibinfo {author} {\bibfnamefont {A.}~\bibnamefont
  {Rudenko}}, \bibinfo {author} {\bibfnamefont {L.}~\bibnamefont {Inhester}},
  \bibinfo {author} {\bibfnamefont {K.}~\bibnamefont {Hanasaki}}, \bibinfo
  {author} {\bibfnamefont {X.}~\bibnamefont {Li}}, \bibinfo {author}
  {\bibfnamefont {S.~J.}\ \bibnamefont {Robatjazi}}, \bibinfo {author}
  {\bibfnamefont {B.}~\bibnamefont {Erk}}, \bibinfo {author} {\bibfnamefont
  {R.}~\bibnamefont {Boll}}, \bibinfo {author} {\bibfnamefont {K.}~\bibnamefont
  {Toyota}}, \bibinfo {author} {\bibfnamefont {Y.}~\bibnamefont {Hao}},
  \bibinfo {author} {\bibfnamefont {O.}~\bibnamefont {Vendrell}}, \emph
  {et~al.},\ }\bibfield  {title} {\bibinfo {title} {Femtosecond response of
  polyatomic molecules to ultra-intense hard {X}-rays},\ }\href
  {https://doi.org/10.1038/nature22373} {\bibfield  {journal} {\bibinfo
  {journal} {Nature}\ }\textbf {\bibinfo {volume} {546}},\ \bibinfo {pages}
  {129} (\bibinfo {year} {2017})}\BibitemShut {NoStop}%
\bibitem [{\citenamefont {Young}\ \emph {et~al.}(2018)\citenamefont {Young},
  \citenamefont {Ueda}, \citenamefont {G\"{u}hr}, \citenamefont {Bucksbaum},
  \citenamefont {Simon}, \citenamefont {Mukamel}, \citenamefont {Rohringer},
  \citenamefont {Prince}, \citenamefont {Masciovecchio}, \citenamefont {Meyer}
  \emph {et~al.}}]{young_roadmap_2018}%
  \BibitemOpen
  \bibfield  {author} {\bibinfo {author} {\bibfnamefont {L.}~\bibnamefont
  {Young}}, \bibinfo {author} {\bibfnamefont {K.}~\bibnamefont {Ueda}},
  \bibinfo {author} {\bibfnamefont {M.}~\bibnamefont {G\"{u}hr}}, \bibinfo
  {author} {\bibfnamefont {P.~H.}\ \bibnamefont {Bucksbaum}}, \bibinfo {author}
  {\bibfnamefont {M.}~\bibnamefont {Simon}}, \bibinfo {author} {\bibfnamefont
  {S.}~\bibnamefont {Mukamel}}, \bibinfo {author} {\bibfnamefont
  {N.}~\bibnamefont {Rohringer}}, \bibinfo {author} {\bibfnamefont {K.~C.}\
  \bibnamefont {Prince}}, \bibinfo {author} {\bibfnamefont {C.}~\bibnamefont
  {Masciovecchio}}, \bibinfo {author} {\bibfnamefont {M.}~\bibnamefont
  {Meyer}}, \emph {et~al.},\ }\bibfield  {title} {\bibinfo {title} {Roadmap of
  ultrafast x-ray atomic and molecular physics},\ }\href
  {https://doi.org/10.1088/1361-6455/aa9735} {\bibfield  {journal} {\bibinfo
  {journal} {J. Phys. B: At. Mol. Opt. Phys.}\ }\textbf {\bibinfo {volume}
  {51}},\ \bibinfo {pages} {032003} (\bibinfo {year} {2018})}\BibitemShut
  {NoStop}%
\bibitem [{\citenamefont {Lindroth}\ \emph {et~al.}(2019)\citenamefont
  {Lindroth}, \citenamefont {Calegari}, \citenamefont {Young}, \citenamefont
  {Harmand}, \citenamefont {Dudovich}, \citenamefont {Berrah},\ and\
  \citenamefont {Smirnova}}]{lindroth_challenges_2019}%
  \BibitemOpen
  \bibfield  {author} {\bibinfo {author} {\bibfnamefont {E.}~\bibnamefont
  {Lindroth}}, \bibinfo {author} {\bibfnamefont {F.}~\bibnamefont {Calegari}},
  \bibinfo {author} {\bibfnamefont {L.}~\bibnamefont {Young}}, \bibinfo
  {author} {\bibfnamefont {M.}~\bibnamefont {Harmand}}, \bibinfo {author}
  {\bibfnamefont {N.}~\bibnamefont {Dudovich}}, \bibinfo {author}
  {\bibfnamefont {N.}~\bibnamefont {Berrah}},\ and\ \bibinfo {author}
  {\bibfnamefont {O.}~\bibnamefont {Smirnova}},\ }\bibfield  {title} {\bibinfo
  {title} {Challenges and opportunities in attosecond and {XFEL} science},\
  }\href {https://doi.org/10.1038/s42254-019-0023-9} {\bibfield  {journal}
  {\bibinfo  {journal} {Nat Rev Phys}\ }\textbf {\bibinfo {volume} {1}},\
  \bibinfo {pages} {107} (\bibinfo {year} {2019})}\BibitemShut {NoStop}%
\bibitem [{\citenamefont {Duris}\ \emph {et~al.}(2020)\citenamefont {Duris},
  \citenamefont {Li}, \citenamefont {Driver}, \citenamefont {Champenois},
  \citenamefont {MacArthur}, \citenamefont {Lutman}, \citenamefont {Zhang},
  \citenamefont {Rosenberger}, \citenamefont {Aldrich}, \citenamefont {Coffee}
  \emph {et~al.}}]{duris_tunable_2020}%
  \BibitemOpen
  \bibfield  {author} {\bibinfo {author} {\bibfnamefont {J.}~\bibnamefont
  {Duris}}, \bibinfo {author} {\bibfnamefont {S.}~\bibnamefont {Li}}, \bibinfo
  {author} {\bibfnamefont {T.}~\bibnamefont {Driver}}, \bibinfo {author}
  {\bibfnamefont {E.~G.}\ \bibnamefont {Champenois}}, \bibinfo {author}
  {\bibfnamefont {J.~P.}\ \bibnamefont {MacArthur}}, \bibinfo {author}
  {\bibfnamefont {A.~A.}\ \bibnamefont {Lutman}}, \bibinfo {author}
  {\bibfnamefont {Z.}~\bibnamefont {Zhang}}, \bibinfo {author} {\bibfnamefont
  {P.}~\bibnamefont {Rosenberger}}, \bibinfo {author} {\bibfnamefont {J.~W.}\
  \bibnamefont {Aldrich}}, \bibinfo {author} {\bibfnamefont {R.}~\bibnamefont
  {Coffee}}, \emph {et~al.},\ }\bibfield  {title} {\bibinfo {title} {Tunable
  isolated attosecond {X}-ray pulses with gigawatt peak power from a
  free-electron laser},\ }\href {https://doi.org/10.1038/s41566-019-0549-5}
  {\bibfield  {journal} {\bibinfo  {journal} {Nat. Photonics}\ }\textbf
  {\bibinfo {volume} {14}},\ \bibinfo {pages} {30} (\bibinfo {year}
  {2020})}\BibitemShut {NoStop}%
\bibitem [{\citenamefont {Li}\ \emph {et~al.}(2022)\citenamefont {Li},
  \citenamefont {Driver}, \citenamefont {Rosenberger}, \citenamefont
  {Champenois}, \citenamefont {Duris}, \citenamefont {Al-Haddad}, \citenamefont
  {Averbukh}, \citenamefont {Barnard}, \citenamefont {Berrah}, \citenamefont
  {Bostedt} \emph {et~al.}}]{li_attosecond_2022}%
  \BibitemOpen
  \bibfield  {author} {\bibinfo {author} {\bibfnamefont {S.}~\bibnamefont
  {Li}}, \bibinfo {author} {\bibfnamefont {T.}~\bibnamefont {Driver}}, \bibinfo
  {author} {\bibfnamefont {P.}~\bibnamefont {Rosenberger}}, \bibinfo {author}
  {\bibfnamefont {E.~G.}\ \bibnamefont {Champenois}}, \bibinfo {author}
  {\bibfnamefont {J.}~\bibnamefont {Duris}}, \bibinfo {author} {\bibfnamefont
  {A.}~\bibnamefont {Al-Haddad}}, \bibinfo {author} {\bibfnamefont
  {V.}~\bibnamefont {Averbukh}}, \bibinfo {author} {\bibfnamefont {J.~C.~T.}\
  \bibnamefont {Barnard}}, \bibinfo {author} {\bibfnamefont {N.}~\bibnamefont
  {Berrah}}, \bibinfo {author} {\bibfnamefont {C.}~\bibnamefont {Bostedt}},
  \emph {et~al.},\ }\bibfield  {title} {\bibinfo {title} {Attosecond coherent
  electron motion in {Auger}-{Meitner} decay},\ }\href
  {https://doi.org/10.1126/science.abj2096} {\bibfield  {journal} {\bibinfo
  {journal} {Science}\ }\textbf {\bibinfo {volume} {375}},\ \bibinfo {pages}
  {285} (\bibinfo {year} {2022})}\BibitemShut {NoStop}%
\bibitem [{\citenamefont {Maroju}\ \emph {et~al.}(2020)\citenamefont {Maroju},
  \citenamefont {Grazioli}, \citenamefont {Di~Fraia}, \citenamefont {Moioli},
  \citenamefont {Ertel}, \citenamefont {Ahmadi}, \citenamefont {Plekan},
  \citenamefont {Finetti}, \citenamefont {Allaria}, \citenamefont {Giannessi}
  \emph {et~al.}}]{maroju_attosecond_2020}%
  \BibitemOpen
  \bibfield  {author} {\bibinfo {author} {\bibfnamefont {P.~K.}\ \bibnamefont
  {Maroju}}, \bibinfo {author} {\bibfnamefont {C.}~\bibnamefont {Grazioli}},
  \bibinfo {author} {\bibfnamefont {M.}~\bibnamefont {Di~Fraia}}, \bibinfo
  {author} {\bibfnamefont {M.}~\bibnamefont {Moioli}}, \bibinfo {author}
  {\bibfnamefont {D.}~\bibnamefont {Ertel}}, \bibinfo {author} {\bibfnamefont
  {H.}~\bibnamefont {Ahmadi}}, \bibinfo {author} {\bibfnamefont
  {O.}~\bibnamefont {Plekan}}, \bibinfo {author} {\bibfnamefont
  {P.}~\bibnamefont {Finetti}}, \bibinfo {author} {\bibfnamefont
  {E.}~\bibnamefont {Allaria}}, \bibinfo {author} {\bibfnamefont
  {L.}~\bibnamefont {Giannessi}}, \emph {et~al.},\ }\bibfield  {title}
  {\bibinfo {title} {Attosecond pulse shaping using a seeded free-electron
  laser},\ }\href {https://doi.org/10.1038/s41586-020-2005-6} {\bibfield
  {journal} {\bibinfo  {journal} {Nature}\ }\textbf {\bibinfo {volume} {578}},\
  \bibinfo {pages} {386} (\bibinfo {year} {2020})}\BibitemShut {NoStop}%
\bibitem [{\citenamefont {Maroju}\ \emph {et~al.}(2023)\citenamefont {Maroju},
  \citenamefont {Di~Fraia}, \citenamefont {Plekan}, \citenamefont {Bonanomi},
  \citenamefont {Merzuk}, \citenamefont {Busto}, \citenamefont {Makos},
  \citenamefont {Schmoll}, \citenamefont {Shah}, \citenamefont {Ribi\v{c}}
  \emph {et~al.}}]{maroju_attosecond_2023}%
  \BibitemOpen
  \bibfield  {author} {\bibinfo {author} {\bibfnamefont {P.~K.}\ \bibnamefont
  {Maroju}}, \bibinfo {author} {\bibfnamefont {M.}~\bibnamefont {Di~Fraia}},
  \bibinfo {author} {\bibfnamefont {O.}~\bibnamefont {Plekan}}, \bibinfo
  {author} {\bibfnamefont {M.}~\bibnamefont {Bonanomi}}, \bibinfo {author}
  {\bibfnamefont {B.}~\bibnamefont {Merzuk}}, \bibinfo {author} {\bibfnamefont
  {D.}~\bibnamefont {Busto}}, \bibinfo {author} {\bibfnamefont
  {I.}~\bibnamefont {Makos}}, \bibinfo {author} {\bibfnamefont
  {M.}~\bibnamefont {Schmoll}}, \bibinfo {author} {\bibfnamefont
  {R.}~\bibnamefont {Shah}}, \bibinfo {author} {\bibfnamefont {P.~R.}\
  \bibnamefont {Ribi\v{c}}}, \emph {et~al.},\ }\bibfield  {title} {\bibinfo
  {title} {Attosecond coherent control of electronic wave packets in two-colour
  photoionization using a novel timing tool for seeded free-electron laser},\
  }\href {https://doi.org/10.1038/s41566-022-01127-3} {\bibfield  {journal}
  {\bibinfo  {journal} {Nat. Photon.}\ }\textbf {\bibinfo {volume} {17}},\
  \bibinfo {pages} {200} (\bibinfo {year} {2023})}\BibitemShut {NoStop}%
\bibitem [{\citenamefont {Lewenstein}\ \emph {et~al.}(1994)\citenamefont
  {Lewenstein}, \citenamefont {Balcou}, \citenamefont {Ivanov}, \citenamefont
  {L'Huillier},\ and\ \citenamefont {Corkum}}]{lewenstein_theory_1994}%
  \BibitemOpen
  \bibfield  {author} {\bibinfo {author} {\bibfnamefont {M.}~\bibnamefont
  {Lewenstein}}, \bibinfo {author} {\bibfnamefont {P.}~\bibnamefont {Balcou}},
  \bibinfo {author} {\bibfnamefont {M.~Y.}\ \bibnamefont {Ivanov}}, \bibinfo
  {author} {\bibfnamefont {A.}~\bibnamefont {L'Huillier}},\ and\ \bibinfo
  {author} {\bibfnamefont {P.~B.}\ \bibnamefont {Corkum}},\ }\bibfield  {title}
  {\bibinfo {title} {Theory of high-harmonic generation by low-frequency laser
  fields},\ }\href {https://doi.org/10.1103/PhysRevA.49.2117} {\bibfield
  {journal} {\bibinfo  {journal} {Phys. Rev. A}\ }\textbf {\bibinfo {volume}
  {49}},\ \bibinfo {pages} {2117} (\bibinfo {year} {1994})}\BibitemShut
  {NoStop}%
\bibitem [{\citenamefont {Nandi}\ \emph {et~al.}(2022)\citenamefont {Nandi},
  \citenamefont {Olofsson}, \citenamefont {Bertolino}, \citenamefont
  {Carlstr{\"o}m}, \citenamefont {Zapata}, \citenamefont {Busto}, \citenamefont
  {Callegari}, \citenamefont {Di~Fraia}, \citenamefont {Eng-Johnsson},
  \citenamefont {Feifel} \emph {et~al.}}]{Nandi2022}%
  \BibitemOpen
  \bibfield  {author} {\bibinfo {author} {\bibfnamefont {S.}~\bibnamefont
  {Nandi}}, \bibinfo {author} {\bibfnamefont {E.}~\bibnamefont {Olofsson}},
  \bibinfo {author} {\bibfnamefont {M.}~\bibnamefont {Bertolino}}, \bibinfo
  {author} {\bibfnamefont {S.}~\bibnamefont {Carlstr{\"o}m}}, \bibinfo {author}
  {\bibfnamefont {F.}~\bibnamefont {Zapata}}, \bibinfo {author} {\bibfnamefont
  {D.}~\bibnamefont {Busto}}, \bibinfo {author} {\bibfnamefont
  {C.}~\bibnamefont {Callegari}}, \bibinfo {author} {\bibfnamefont
  {M.}~\bibnamefont {Di~Fraia}}, \bibinfo {author} {\bibfnamefont
  {P.}~\bibnamefont {Eng-Johnsson}}, \bibinfo {author} {\bibfnamefont
  {R.}~\bibnamefont {Feifel}}, \emph {et~al.},\ }\bibfield  {title} {\bibinfo
  {title} {Observation of rabi dynamics with a short-wavelength free-electron
  laser},\ }\href {https://doi.org/10.1038/s41586-022-04948-y} {\bibfield
  {journal} {\bibinfo  {journal} {Nature}\ }\textbf {\bibinfo {volume} {608}},\
  \bibinfo {pages} {488} (\bibinfo {year} {2022})}\BibitemShut {NoStop}%
\bibitem [{\citenamefont {Rabi}(1937)}]{Rabi1937}%
  \BibitemOpen
  \bibfield  {author} {\bibinfo {author} {\bibfnamefont {I.~I.}\ \bibnamefont
  {Rabi}},\ }\bibfield  {title} {\bibinfo {title} {Space quantization in a
  gyrating magnetic field},\ }\href {https://doi.org/10.1103/PhysRev.51.652}
  {\bibfield  {journal} {\bibinfo  {journal} {Phys. Rev.}\ }\textbf {\bibinfo
  {volume} {51}},\ \bibinfo {pages} {652} (\bibinfo {year} {1937})}\BibitemShut
  {NoStop}%
\bibitem [{\citenamefont {Autler}\ and\ \citenamefont
  {Townes}(1955)}]{Autler1955}%
  \BibitemOpen
  \bibfield  {author} {\bibinfo {author} {\bibfnamefont {S.~H.}\ \bibnamefont
  {Autler}}\ and\ \bibinfo {author} {\bibfnamefont {C.~H.}\ \bibnamefont
  {Townes}},\ }\bibfield  {title} {\bibinfo {title} {Stark effect in rapidly
  varying fields},\ }\href {https://doi.org/10.1103/PhysRev.100.703} {\bibfield
   {journal} {\bibinfo  {journal} {Phys. Rev.}\ }\textbf {\bibinfo {volume}
  {100}},\ \bibinfo {pages} {703} (\bibinfo {year} {1955})}\BibitemShut
  {NoStop}%
\bibitem [{\citenamefont {Rodríguez}\ \emph {et~al.}(2009)\citenamefont
  {Rodríguez}, \citenamefont {Macri},\ and\ \citenamefont
  {Arbó}}]{rodriguez_resonant-enhanced_2009}%
  \BibitemOpen
  \bibfield  {author} {\bibinfo {author} {\bibfnamefont {V.~D.}\ \bibnamefont
  {Rodríguez}}, \bibinfo {author} {\bibfnamefont {P.~A.}\ \bibnamefont
  {Macri}},\ and\ \bibinfo {author} {\bibfnamefont {D.~G.}\ \bibnamefont
  {Arbó}},\ }\bibfield  {title} {\bibinfo {title} {Resonant-enhanced
  above-threshold ionization of atoms by {XUV} short laser pulses},\ }\href
  {https://doi.org/10.1016/j.nimb.2008.10.066} {\bibfield  {journal} {\bibinfo
  {journal} {Nuclear Instruments and Methods in Physics Research Section B:
  Beam Interactions with Materials and Atoms}\ }\bibinfo {series} {Proceedings
  of the {Fourth} {International} {Conference} on {Elementary} {Processes} in
  {Atomic} {Systems}},\ \textbf {\bibinfo {volume} {267}},\ \bibinfo {pages}
  {334} (\bibinfo {year} {2009})}\BibitemShut {NoStop}%
\bibitem [{\citenamefont {T\'oth}\ and\ \citenamefont
  {Csehi}(2021)}]{Toth2021}%
  \BibitemOpen
  \bibfield  {author} {\bibinfo {author} {\bibfnamefont {A.}~\bibnamefont
  {T\'oth}}\ and\ \bibinfo {author} {\bibfnamefont {A.}~\bibnamefont {Csehi}},\
  }\bibfield  {title} {\bibinfo {title} {Probing strong-field two-photon
  transitions through dynamic interference},\ }\href
  {https://doi.org/10.1088/1361-6455/abdb8e} {\bibfield  {journal} {\bibinfo
  {journal} {Journal of Physics B: Atomic, Molecular and Optical Physics}\
  }\textbf {\bibinfo {volume} {54}},\ \bibinfo {pages} {035005} (\bibinfo
  {year} {2021})}\BibitemShut {NoStop}%
\bibitem [{\citenamefont {Younis}\ and\ \citenamefont
  {Eberly}(2022)}]{Younis2022}%
  \BibitemOpen
  \bibfield  {author} {\bibinfo {author} {\bibfnamefont {D.}~\bibnamefont
  {Younis}}\ and\ \bibinfo {author} {\bibfnamefont {J.~H.}\ \bibnamefont
  {Eberly}},\ }\bibfield  {title} {\bibinfo {title} {Benchmark of few-level
  quantum theory vs ab initio numerical solutions for the strong-field
  autler–townes effect in photoionization of hydrogen},\ }\href
  {https://doi.org/10.1088/1361-6455/ac7d7f} {\bibfield  {journal} {\bibinfo
  {journal} {Journal of Physics B: Atomic, Molecular and Optical Physics}\
  }\textbf {\bibinfo {volume} {55}},\ \bibinfo {pages} {164001} (\bibinfo
  {year} {2022})}\BibitemShut {NoStop}%
\bibitem [{\citenamefont {Bunjac}\ \emph {et~al.}(2022)\citenamefont {Bunjac},
  \citenamefont {Popovi\'c},\ and\ \citenamefont {Simonovi\'c}}]{Bunjac2022}%
  \BibitemOpen
  \bibfield  {author} {\bibinfo {author} {\bibfnamefont {A.}~\bibnamefont
  {Bunjac}}, \bibinfo {author} {\bibfnamefont {D.~B.}\ \bibnamefont
  {Popovi\'c}},\ and\ \bibinfo {author} {\bibfnamefont {N.~S.}\ \bibnamefont
  {Simonovi\'c}},\ }\bibfield  {title} {\bibinfo {title} {Analysis of the
  asymmetry of autler–townes doublets in the energy spectra of photoelectrons
  produced at two-photon ionization of atoms by strong laser pulses},\ }\href
  {https://doi.org/10.1140/epjd/s10053-022-00572-7} {\bibfield  {journal}
  {\bibinfo  {journal} {The European Physical Journal D}\ }\textbf {\bibinfo
  {volume} {76}},\ \bibinfo {pages} {249} (\bibinfo {year} {2022})}\BibitemShut
  {NoStop}%
\bibitem [{\citenamefont {Zhang}\ \emph {et~al.}(2022)\citenamefont {Zhang},
  \citenamefont {Zhou}, \citenamefont {Liao}, \citenamefont {Chen},
  \citenamefont {Liang}, \citenamefont {Ke}, \citenamefont {Li}, \citenamefont
  {Csehi},\ and\ \citenamefont {Lu}}]{Zhang2022}%
  \BibitemOpen
  \bibfield  {author} {\bibinfo {author} {\bibfnamefont {X.}~\bibnamefont
  {Zhang}}, \bibinfo {author} {\bibfnamefont {Y.}~\bibnamefont {Zhou}},
  \bibinfo {author} {\bibfnamefont {Y.}~\bibnamefont {Liao}}, \bibinfo {author}
  {\bibfnamefont {Y.}~\bibnamefont {Chen}}, \bibinfo {author} {\bibfnamefont
  {J.}~\bibnamefont {Liang}}, \bibinfo {author} {\bibfnamefont
  {Q.}~\bibnamefont {Ke}}, \bibinfo {author} {\bibfnamefont {M.}~\bibnamefont
  {Li}}, \bibinfo {author} {\bibfnamefont {A.}~\bibnamefont {Csehi}},\ and\
  \bibinfo {author} {\bibfnamefont {P.}~\bibnamefont {Lu}},\ }\bibfield
  {title} {\bibinfo {title} {Effect of nonresonant states in near-resonant
  two-photon ionization of hydrogen},\ }\href
  {https://doi.org/10.1103/PhysRevA.106.063114} {\bibfield  {journal} {\bibinfo
   {journal} {Phys. Rev. A}\ }\textbf {\bibinfo {volume} {106}},\ \bibinfo
  {pages} {063114} (\bibinfo {year} {2022})}\BibitemShut {NoStop}%
\bibitem [{\citenamefont {LaGattuta}(1993)}]{LaGattuta1993}%
  \BibitemOpen
  \bibfield  {author} {\bibinfo {author} {\bibfnamefont {K.~J.}\ \bibnamefont
  {LaGattuta}},\ }\bibfield  {title} {\bibinfo {title} {Above-threshold
  ionization of atomic hydrogen via resonant intermediate states},\ }\href
  {https://doi.org/10.1103/PhysRevA.47.1560} {\bibfield  {journal} {\bibinfo
  {journal} {Phys. Rev. A}\ }\textbf {\bibinfo {volume} {47}},\ \bibinfo
  {pages} {1560} (\bibinfo {year} {1993})}\BibitemShut {NoStop}%
\bibitem [{\citenamefont {Girju}\ \emph {et~al.}(2007)\citenamefont {Girju},
  \citenamefont {Hristov}, \citenamefont {Kidun},\ and\ \citenamefont
  {Bauer}}]{Girju2007}%
  \BibitemOpen
  \bibfield  {author} {\bibinfo {author} {\bibfnamefont {M.~G.}\ \bibnamefont
  {Girju}}, \bibinfo {author} {\bibfnamefont {K.}~\bibnamefont {Hristov}},
  \bibinfo {author} {\bibfnamefont {O.}~\bibnamefont {Kidun}},\ and\ \bibinfo
  {author} {\bibfnamefont {D.}~\bibnamefont {Bauer}},\ }\bibfield  {title}
  {\bibinfo {title} {Nonperturbative resonant strong field ionization of atomic
  hydrogen},\ }\href {https://doi.org/10.1088/0953-4075/40/21/004} {\bibfield
  {journal} {\bibinfo  {journal} {Journal of Physics B: Atomic, Molecular and
  Optical Physics}\ }\textbf {\bibinfo {volume} {40}},\ \bibinfo {pages} {4165}
  (\bibinfo {year} {2007})}\BibitemShut {NoStop}%
\bibitem [{\citenamefont {Beers}\ and\ \citenamefont
  {Armstrong}(1975)}]{Beers1975}%
  \BibitemOpen
  \bibfield  {author} {\bibinfo {author} {\bibfnamefont {B.~L.}\ \bibnamefont
  {Beers}}\ and\ \bibinfo {author} {\bibfnamefont {L.}~\bibnamefont
  {Armstrong}},\ }\bibfield  {title} {\bibinfo {title} {Exact solution of a
  realistic model for two-photon ionization},\ }\href
  {https://doi.org/10.1103/PhysRevA.12.2447} {\bibfield  {journal} {\bibinfo
  {journal} {Phys. Rev. A}\ }\textbf {\bibinfo {volume} {12}},\ \bibinfo
  {pages} {2447} (\bibinfo {year} {1975})}\BibitemShut {NoStop}%
\bibitem [{\citenamefont {Knight}(1977)}]{knight_saturation_1977}%
  \BibitemOpen
  \bibfield  {author} {\bibinfo {author} {\bibfnamefont {P.~L.}\ \bibnamefont
  {Knight}},\ }\bibfield  {title} {\bibinfo {title} {Saturation and rabi
  oscillations in resonant multiphoton ionization},\ }\href
  {https://doi.org/10.1016/0030-4018(77)90013-X} {\bibfield  {journal}
  {\bibinfo  {journal} {Optics Communications}\ }\textbf {\bibinfo {volume}
  {22}},\ \bibinfo {pages} {173} (\bibinfo {year} {1977})}\BibitemShut
  {NoStop}%
\bibitem [{\citenamefont {Holt}\ \emph {et~al.}(1983)\citenamefont {Holt},
  \citenamefont {Raymer},\ and\ \citenamefont {Reinhardt}}]{Holt1983}%
  \BibitemOpen
  \bibfield  {author} {\bibinfo {author} {\bibfnamefont {C.~R.}\ \bibnamefont
  {Holt}}, \bibinfo {author} {\bibfnamefont {M.~G.}\ \bibnamefont {Raymer}},\
  and\ \bibinfo {author} {\bibfnamefont {W.~P.}\ \bibnamefont {Reinhardt}},\
  }\bibfield  {title} {\bibinfo {title} {Time dependences of two-, three-, and
  four-photon ionization of atomic hydrogen in the ground $1{}^2{S}$ and
  metastable $2{}^2{S}$ states},\ }\href
  {https://doi.org/10.1103/PhysRevA.27.2971} {\bibfield  {journal} {\bibinfo
  {journal} {Phys. Rev. A}\ }\textbf {\bibinfo {volume} {27}},\ \bibinfo
  {pages} {2971} (\bibinfo {year} {1983})}\BibitemShut {NoStop}%
\bibitem [{\citenamefont {Cohen-Tannoudji}\ \emph {et~al.}(1998)\citenamefont
  {Cohen-Tannoudji}, \citenamefont {Dupont-Roc},\ and\ \citenamefont
  {Grynberg}}]{CT1998}%
  \BibitemOpen
  \bibfield  {author} {\bibinfo {author} {\bibfnamefont {C.}~\bibnamefont
  {Cohen-Tannoudji}}, \bibinfo {author} {\bibfnamefont {J.}~\bibnamefont
  {Dupont-Roc}},\ and\ \bibinfo {author} {\bibfnamefont {G.}~\bibnamefont
  {Grynberg}},\ }\bibinfo {title} {Nonperturbative calculation of transition
  amplitudes},\ in\ \href
  {https://doi.org/https://doi.org/10.1002/9783527617197.ch3} {\emph {\bibinfo
  {booktitle} {Atom—Photon Interactions}}}\ (\bibinfo  {publisher} {John
  Wiley \& Sons, Ltd},\ \bibinfo {year} {1998})\ Chap.~\bibinfo {chapter} {3},
  pp.\ \bibinfo {pages} {165--255}\BibitemShut {NoStop}%
\bibitem [{\citenamefont {Palacios}\ \emph {et~al.}(2006)\citenamefont
  {Palacios}, \citenamefont {Bachau},\ and\ \citenamefont
  {Martín}}]{palacios_step-ladder_2006}%
  \BibitemOpen
  \bibfield  {author} {\bibinfo {author} {\bibfnamefont {A.}~\bibnamefont
  {Palacios}}, \bibinfo {author} {\bibfnamefont {H.}~\bibnamefont {Bachau}},\
  and\ \bibinfo {author} {\bibfnamefont {F.}~\bibnamefont {Martín}},\
  }\bibfield  {title} {\bibinfo {title} {Step-ladder {Rabi} oscillations in
  molecules exposed to intense ultrashort vuv pulses},\ }\href
  {https://doi.org/10.1103/PhysRevA.74.031402} {\bibfield  {journal} {\bibinfo
  {journal} {Phys. Rev. A}\ }\textbf {\bibinfo {volume} {74}},\ \bibinfo
  {pages} {031402} (\bibinfo {year} {2006})}\BibitemShut {NoStop}%
\bibitem [{\citenamefont {Demekhin}\ and\ \citenamefont
  {Cederbaum}(2012)}]{Demekhin2012}%
  \BibitemOpen
  \bibfield  {author} {\bibinfo {author} {\bibfnamefont {P.~V.}\ \bibnamefont
  {Demekhin}}\ and\ \bibinfo {author} {\bibfnamefont {L.~S.}\ \bibnamefont
  {Cederbaum}},\ }\bibfield  {title} {\bibinfo {title} {Coherent intense
  resonant laser pulses lead to interference in the time domain seen in the
  spectrum of the emitted particles},\ }\href
  {https://doi.org/10.1103/PhysRevA.86.063412} {\bibfield  {journal} {\bibinfo
  {journal} {Phys. Rev. A}\ }\textbf {\bibinfo {volume} {86}},\ \bibinfo
  {pages} {063412} (\bibinfo {year} {2012})}\BibitemShut {NoStop}%
\bibitem [{\citenamefont {Greenman}\ \emph {et~al.}(2010)\citenamefont
  {Greenman}, \citenamefont {Ho}, \citenamefont {Pabst}, \citenamefont
  {Kamarchik}, \citenamefont {Mazziotti},\ and\ \citenamefont
  {Santra}}]{greenmanImplementationPRA2010}%
  \BibitemOpen
  \bibfield  {author} {\bibinfo {author} {\bibfnamefont {L.}~\bibnamefont
  {Greenman}}, \bibinfo {author} {\bibfnamefont {P.~J.}\ \bibnamefont {Ho}},
  \bibinfo {author} {\bibfnamefont {S.}~\bibnamefont {Pabst}}, \bibinfo
  {author} {\bibfnamefont {E.}~\bibnamefont {Kamarchik}}, \bibinfo {author}
  {\bibfnamefont {D.~A.}\ \bibnamefont {Mazziotti}},\ and\ \bibinfo {author}
  {\bibfnamefont {R.}~\bibnamefont {Santra}},\ }\bibfield  {title} {\bibinfo
  {title} {Implementation of the time-dependent configuration-interaction
  singles method for atomic strong-field processes},\ }\href
  {https://doi.org/10.1103/PhysRevA.82.023406} {\bibfield  {journal} {\bibinfo
  {journal} {Phys. Rev. A}\ }\textbf {\bibinfo {volume} {82}},\ \bibinfo
  {pages} {023406} (\bibinfo {year} {2010})}\BibitemShut {NoStop}%
\bibitem [{\citenamefont {Simon}(1979)}]{Simon1979}%
  \BibitemOpen
  \bibfield  {author} {\bibinfo {author} {\bibfnamefont {B.}~\bibnamefont
  {Simon}},\ }\bibfield  {title} {\bibinfo {title} {The definition of molecular
  resonance curves by the method of exterior complex scaling},\ }\href
  {https://doi.org/https://doi.org/10.1016/0375-9601(79)90165-8} {\bibfield
  {journal} {\bibinfo  {journal} {Physics Letters A}\ }\textbf {\bibinfo
  {volume} {71}},\ \bibinfo {pages} {211} (\bibinfo {year} {1979})}\BibitemShut
  {NoStop}%
\bibitem [{\citenamefont {Moiseyev}(1998)}]{Moiseyev1998}%
  \BibitemOpen
  \bibfield  {author} {\bibinfo {author} {\bibfnamefont {N.}~\bibnamefont
  {Moiseyev}},\ }\bibfield  {title} {\bibinfo {title} {Quantum theory of
  resonances: calculating energies, widths and cross-sections by complex
  scaling},\ }\href
  {https://doi.org/https://doi.org/10.1016/S0370-1573(98)00002-7} {\bibfield
  {journal} {\bibinfo  {journal} {Physics Reports}\ }\textbf {\bibinfo {volume}
  {302}},\ \bibinfo {pages} {212} (\bibinfo {year} {1998})}\BibitemShut
  {NoStop}%
\bibitem [{\citenamefont {Cormier}\ and\ \citenamefont
  {Lambropoulos}(1995)}]{cormier_extrapolation_1995}%
  \BibitemOpen
  \bibfield  {author} {\bibinfo {author} {\bibfnamefont {E.}~\bibnamefont
  {Cormier}}\ and\ \bibinfo {author} {\bibfnamefont {P.}~\bibnamefont
  {Lambropoulos}},\ }\bibfield  {title} {\bibinfo {title} {Extrapolation method
  for the evaluation of above threshold ionization cross sections for one- and
  two-electron systems},\ }\href {https://doi.org/10.1088/0953-4075/28/23/013}
  {\bibfield  {journal} {\bibinfo  {journal} {J. Phys. B: At. Mol. Opt. Phys.}\
  }\textbf {\bibinfo {volume} {28}},\ \bibinfo {pages} {5043} (\bibinfo {year}
  {1995})}\BibitemShut {NoStop}%
\bibitem [{\citenamefont {Rogus}\ and\ \citenamefont
  {Lewenstein}(1986)}]{Rogus1986}%
  \BibitemOpen
  \bibfield  {author} {\bibinfo {author} {\bibfnamefont {R.}~\bibnamefont
  {Rogus}}\ and\ \bibinfo {author} {\bibfnamefont {M.}~\bibnamefont
  {Lewenstein}},\ }\bibfield  {title} {\bibinfo {title} {Resonant ionisation by
  smooth laser pulses},\ }\href {https://doi.org/10.1088/0022-3700/19/19/018}
  {\bibfield  {journal} {\bibinfo  {journal} {J. Phys. B: Atom. Mol. Phys.}\
  }\textbf {\bibinfo {volume} {19}},\ \bibinfo {pages} {3051} (\bibinfo {year}
  {1986})}\BibitemShut {NoStop}%
\bibitem [{\citenamefont {Simonovi\'c}\ \emph {et~al.}(2023)\citenamefont
  {Simonovi\'c}, \citenamefont {Popovi\'c},\ and\ \citenamefont
  {Bunjac}}]{simonovic_manifestations_2023}%
  \BibitemOpen
  \bibfield  {author} {\bibinfo {author} {\bibfnamefont {N.~S.}\ \bibnamefont
  {Simonovi\'c}}, \bibinfo {author} {\bibfnamefont {D.~B.}\ \bibnamefont
  {Popovi\'c}},\ and\ \bibinfo {author} {\bibfnamefont {A.}~\bibnamefont
  {Bunjac}},\ }\bibfield  {title} {\bibinfo {title} {Manifestations of {Rabi}
  {Dynamics} in the {Photoelectron} {Energy} {Spectra} at {Resonant}
  {Two}-{Photon} {Ionization} of {Atom} by {Intense} {Short} {Laser}
  {Pulses}},\ }\href {https://doi.org/10.3390/atoms11020020} {\bibfield
  {journal} {\bibinfo  {journal} {Atoms}\ }\textbf {\bibinfo {volume} {11}},\
  \bibinfo {pages} {20} (\bibinfo {year} {2023})}\BibitemShut {NoStop}%
\bibitem [{\citenamefont {Harkema}\ \emph {et~al.}(2021)\citenamefont
  {Harkema}, \citenamefont {Cariker}, \citenamefont {Lindroth}, \citenamefont
  {Argenti},\ and\ \citenamefont {Sandhu}}]{harkema_autoionizing_2021}%
  \BibitemOpen
  \bibfield  {author} {\bibinfo {author} {\bibfnamefont {N.}~\bibnamefont
  {Harkema}}, \bibinfo {author} {\bibfnamefont {C.}~\bibnamefont {Cariker}},
  \bibinfo {author} {\bibfnamefont {E.}~\bibnamefont {Lindroth}}, \bibinfo
  {author} {\bibfnamefont {L.}~\bibnamefont {Argenti}},\ and\ \bibinfo {author}
  {\bibfnamefont {A.}~\bibnamefont {Sandhu}},\ }\bibfield  {title} {\bibinfo
  {title} {Autoionizing {Polaritons} in {Attosecond} {Atomic} {Ionization}},\
  }\href {https://doi.org/10.1103/PhysRevLett.127.023202} {\bibfield  {journal}
  {\bibinfo  {journal} {Phys. Rev. Lett.}\ }\textbf {\bibinfo {volume} {127}},\
  \bibinfo {pages} {023202} (\bibinfo {year} {2021})}\BibitemShut {NoStop}%
\bibitem [{\citenamefont {Wollenhaupt}\ \emph {et~al.}(2003)\citenamefont
  {Wollenhaupt}, \citenamefont {Assion}, \citenamefont {Bazhan}, \citenamefont
  {Horn}, \citenamefont {Liese}, \citenamefont {Sarpe-Tudoran}, \citenamefont
  {Winter},\ and\ \citenamefont {Baumert}}]{Wollenhaupt2003}%
  \BibitemOpen
  \bibfield  {author} {\bibinfo {author} {\bibfnamefont {M.}~\bibnamefont
  {Wollenhaupt}}, \bibinfo {author} {\bibfnamefont {A.}~\bibnamefont {Assion}},
  \bibinfo {author} {\bibfnamefont {O.}~\bibnamefont {Bazhan}}, \bibinfo
  {author} {\bibfnamefont {C.}~\bibnamefont {Horn}}, \bibinfo {author}
  {\bibfnamefont {D.}~\bibnamefont {Liese}}, \bibinfo {author} {\bibfnamefont
  {C.}~\bibnamefont {Sarpe-Tudoran}}, \bibinfo {author} {\bibfnamefont
  {M.}~\bibnamefont {Winter}},\ and\ \bibinfo {author} {\bibfnamefont
  {T.}~\bibnamefont {Baumert}},\ }\bibfield  {title} {\bibinfo {title} {Control
  of interferences in an autler-townes doublet: Symmetry of control
  parameters},\ }\href {https://doi.org/10.1103/PhysRevA.68.015401} {\bibfield
  {journal} {\bibinfo  {journal} {Phys. Rev. A}\ }\textbf {\bibinfo {volume}
  {68}},\ \bibinfo {pages} {015401} (\bibinfo {year} {2003})}\BibitemShut
  {NoStop}%
\bibitem [{\citenamefont {Wollenhaupt}\ \emph {et~al.}(2006)\citenamefont
  {Wollenhaupt}, \citenamefont {Pr{\"a}kelt}, \citenamefont {Sarpe-Tudoran},
  \citenamefont {Liese},\ and\ \citenamefont {Baumert}}]{Wollenhaupt2006}%
  \BibitemOpen
  \bibfield  {author} {\bibinfo {author} {\bibfnamefont {M.}~\bibnamefont
  {Wollenhaupt}}, \bibinfo {author} {\bibfnamefont {A.}~\bibnamefont
  {Pr{\"a}kelt}}, \bibinfo {author} {\bibfnamefont {C.}~\bibnamefont
  {Sarpe-Tudoran}}, \bibinfo {author} {\bibfnamefont {D.}~\bibnamefont
  {Liese}},\ and\ \bibinfo {author} {\bibfnamefont {T.}~\bibnamefont
  {Baumert}},\ }\bibfield  {title} {\bibinfo {title} {Quantum control by
  selective population of dressed states using intense chirped femtosecond
  laser pulses},\ }\href {https://doi.org/10.1007/s00340-005-2066-0} {\bibfield
   {journal} {\bibinfo  {journal} {Applied Physics B}\ }\textbf {\bibinfo
  {volume} {82}},\ \bibinfo {pages} {183} (\bibinfo {year} {2006})}\BibitemShut
  {NoStop}%
\bibitem [{\citenamefont {Saalmann}\ \emph {et~al.}(2018)\citenamefont
  {Saalmann}, \citenamefont {Giri},\ and\ \citenamefont {Rost}}]{Saalmann2018}%
  \BibitemOpen
  \bibfield  {author} {\bibinfo {author} {\bibfnamefont {U.}~\bibnamefont
  {Saalmann}}, \bibinfo {author} {\bibfnamefont {S.~K.}\ \bibnamefont {Giri}},\
  and\ \bibinfo {author} {\bibfnamefont {J.~M.}\ \bibnamefont {Rost}},\
  }\bibfield  {title} {\bibinfo {title} {Adiabatic passage to the continuum:
  Controlling ionization with chirped laser pulses},\ }\href
  {https://doi.org/10.1103/PhysRevLett.121.153203} {\bibfield  {journal}
  {\bibinfo  {journal} {Phys. Rev. Lett.}\ }\textbf {\bibinfo {volume} {121}},\
  \bibinfo {pages} {153203} (\bibinfo {year} {2018})}\BibitemShut {NoStop}%
\bibitem [{\citenamefont {Maquet}\ \emph {et~al.}(1983)\citenamefont {Maquet},
  \citenamefont {Chu},\ and\ \citenamefont {Reinhardt}}]{maquet_stark_1983}%
  \BibitemOpen
  \bibfield  {author} {\bibinfo {author} {\bibfnamefont {A.}~\bibnamefont
  {Maquet}}, \bibinfo {author} {\bibfnamefont {S.-I.}\ \bibnamefont {Chu}},\
  and\ \bibinfo {author} {\bibfnamefont {W.~P.}\ \bibnamefont {Reinhardt}},\
  }\bibfield  {title} {\bibinfo {title} {Stark ionization in dc and ac fields:
  {An} ${L}^2$ complex-coordinate approach},\ }\href
  {https://doi.org/10.1103/PhysRevA.27.2946} {\bibfield  {journal} {\bibinfo
  {journal} {Phys. Rev. A}\ }\textbf {\bibinfo {volume} {27}},\ \bibinfo
  {pages} {2946} (\bibinfo {year} {1983})}\BibitemShut {NoStop}%
\bibitem [{\citenamefont {Pahl}\ \emph {et~al.}(1996)\citenamefont {Pahl},
  \citenamefont {Meyer},\ and\ \citenamefont {Cederbaum}}]{Pahl1996}%
  \BibitemOpen
  \bibfield  {author} {\bibinfo {author} {\bibfnamefont {E.}~\bibnamefont
  {Pahl}}, \bibinfo {author} {\bibfnamefont {H.-D.}\ \bibnamefont {Meyer}},\
  and\ \bibinfo {author} {\bibfnamefont {L.~S.}\ \bibnamefont {Cederbaum}},\
  }\bibfield  {title} {\bibinfo {title} {Competition between excitation and
  electronic decay of short-lived molecular states},\ }\href
  {https://doi.org/10.1007/s004600050086} {\bibfield  {journal} {\bibinfo
  {journal} {Zeitschrift f{\"u}r Physik D Atoms, Molecules and Clusters}\
  }\textbf {\bibinfo {volume} {38}},\ \bibinfo {pages} {215} (\bibinfo {year}
  {1996})}\BibitemShut {NoStop}%
\end{thebibliography}%
\end{document}